\documentclass[11pt]{article}
\pdfoutput=1
\usepackage{array}
\usepackage{cite}
\usepackage{calc}
\usepackage{caption}
\usepackage{graphicx}
\usepackage{epstopdf}
\usepackage{psfrag}
\usepackage{subcaption}
\usepackage{stfloats}
\usepackage{longtable}
\usepackage{amsmath,amsfonts,amssymb,amsbsy,bm,paralist,theorem,ifthen,color}
\usepackage{algorithm}
\usepackage{algpseudocode}
\usepackage{adjustbox}
\usepackage{booktabs}

\usepackage[dvipsnames]{xcolor}
\newcommand{\beq}{\begin{equation}}
\newcommand{\eeq}{\end{equation}}

\bigskip

\usepackage{soul}
\usepackage{authblk}

\begin{document}
	\title{Machine Learning to Generate Adjustable Dose Distributions in Head-and-Neck Cancer Radiation Therapy}
	\date{}

\author[1]{Davood Hajinezhad\thanks{davood.hajinezhad@sas.com}}
\author[1]{Afshin Oroojlooy}
\author[1]{Mohammadreza Nazari}
\author[1]{Xin Hunt}
\author[1]{Jorge Silva}
\author[2]{Colette Shen}
\author[2]{Bhisham Chera}
\author[2]{Shiva K. Das}
\affil[1]{SAS Institute Inc., Cary, NC, USA}
\affil[2]{Department of Radiation Oncology, University of North Carolina at Chapel Hill, USA}
\maketitle
	\begin{abstract}
		In this work, we propose a Machine Learning model that generates an adjustable 3D dose distribution for external beam radiation therapy for head-and-neck cancer treatments. In contrast to existing Machine Learning methods that provide a single model, we create pairs of models for each organ-at-risk, namely lower-extreme and upper-extreme models.    These model pairs for an organ-at-risk propose doses that give lower and higher doses to that organ-at-risk, while also encapsulating the dose trade-off to other organs-at-risk.  By weighting and combining the model pairs for all organs-at-risk, we are able to dynamically create adjustable dose distributions that can be used, in real-time, to move doses between organs-at-risk, thereby customizing the dose distribution to the needs of a particular patient. We leverage a key observation that the training data set inherently contains the clinical trade-offs.  We show that the adjustable distributions are able to provide reasonable clinical dose latitude in the trade-off of doses between organs-at-risk. 
	\end{abstract}
	
{\it Keywords:} radiation therapy, IMRT, VMAT, head-and-neck, treatment planning, machine learning
\section{INTRODUCTION}\label{sub:intro} 

Radiation therapy (RT) for cancer treatment is built on the paradigm that cancerous tissue may potentially be eliminated by subjecting it to therapeutic radiation doses.  However, external beam megavoltage radiation therapy, delivered using linear accelerators, consists of radiation beams that necessarily traverse a path through normal healthy tissue to reach the tumor, then exiting through normal tissue beyond the tumor.  Thus, the challenge is to provide sufficient therapeutic doses to cancerous tissues while maximally sparing surrounding healthy organs-at-risk (OARs) from collateral damage. In Head-and-Neck (HN) cancer, delivery of therapeutic target dose is complicated by the presence of several critical normal organs (larynx, parotid glands, spinal cord, brainstem, oral cavity, brain) in close proximity to the tumor target.  In such cases, radiation therapy is delivered via either Intensity Modulated Radiation Therapy (IMRT) \cite{cedric2002clinical} or Volumetric  Modulated Arc Therapy (VMAT) \cite{otto2008volumetric}.  

IMRT/VMAT delivers dose to the target via linear accelerator beams directed from several different orientations. In each orientation, the intensity of the radiation beam exiting the linear accelerator aperture is optimized to a spatially non-uniform distribution across the aperture, so as to reduce doses to critical organs in the path of the beam. Treatment planning systems (TPS) running optimization algorithms are used to create a deliverable plan that modulates the intensity of the beams across every point in the aperture with the goal of reducing dose to OARs (below specified constraint levels) while satisfying target prescription dose coverage \cite{unkelbach2015optimization, christiansen2018continuous,yan2018fast,dursun2019column}.  The problem is complicated by three factors: 
(a) the close proximity of OARs to tumor target and the large number of beam intensity variables; (b) in some cases, based on the geometry, it may not be possible to reduce one or more OAR  doses below constraint levels; and (c) the desire to reduce doses to OARs as much as possible (even beyond the reduction called for by the constraint levels) to minimize patient complications. The process of optimizing a patient’s treatment plan in the TPS is not straightforward. The human planner continuously interacts with the computer optimization algorithm to adjust OAR/target importance weights and constraints to direct the course of the optimization based on the planner’s perception of how the reduction in dose to one OAR negatively affects (i.e., increases) doses to other OARs. Clearly, it is difficult for a human planner to comprehend the full extent of trade-off in doses between multiple OARs, especially in head-and-neck treatment planning.   A consequence is that the resulting treatment plan dose distribution prepared by a human planner will invariably have differences from that prepared by another planner. An additional source of variability is clinician prioritization of dose reduction to one or more OARs over others.  The OARs to prioritize can vary from patient to patient based on patient-specific clinical factors.  

Due to the complexity of the treatment planning process, various Machine Learning (ML) implementations  have sought to generate automatic plans using the database of plans previously created by experienced human planners. The majority of these Machine Learning implementations take the manually contoured OARs/tumor target drawn on the individual CT slices of the 3D dataset  and the therapeutic prescription dose as inputs. The Machine Learning output is a 3D dose distribution that is intended to mimic the distribution that would have been generated by an experienced human planner (in some cases the outputs go so far as to predict the linear accelerator parameters required to generate the dose distribution as well).  In \cite{fan2019automatic}, a residual deep neural network (ResNet) was trained to predict does vales for head-neck cancer. A volumetric dose prediction model, named DoseNet, was proposed in \cite{kearney2018dosenet} to predict dose values. In \cite{barragan2019three}, the beam configurations are added to the model input to generalize the model in the sense that the model does not need to be retrained for the new set of beam configurations. These ML models provide promising solutions for 3D dose distributions in a much shorter amount of time compared to the classical optimization algorithms. See the survey article \cite{momin2021knowledge} for a more comprehensive review.

Utilizing the previously mentioned ML approaches results in a single dose distribution plan for each patient. The plan in its entirety is required to be accepted by the clinician since these ML approaches do not have a way to adjust the dose distribution. If, however, the clinician was not entirely satisfied with the plan and wanted to lower doses to a particular OAR, the only recourse would be to manually redo the plan from scratch with a human planner.  Needless to say, manual replanning is a time consuming task that obviates the ML solution and points to the clinical need for {\it adjustable} ML dose distribution plans that allow trade-off in doses between organs.


To address this clinical need, there are existing approaches that generate Pareto solutions for the dose distribution problem \cite{nguyen2019generating, bohara2020using}. These methods rely on the generation of many Pareto optimal solutions with a variety of weight vectors. To obtain the corresponding training datasets, one needs different tuples of (organ-weight, Pareto-solutions) for each patient, i.e., the input and the labels in the ML model. Obtaining large enough different tuples of (organ-weight, Pareto-solutions) to train accurate ML models is computationally expensive and time-consuming, since it requires solving an optimization problem for each single tuple. These challenges motivated us to propose a different approach.

A key observation we leverage in dealing with these issues is to exploit the notion that the manually generated clinical plans in the training set contain inherent variability, attributable to the randomness associated with the trade-off trajectory selected by the human planners during the planning process as well as to the variability associated with clinician choice to increase sparing to selective OARs based on patient-specific clinical factors. This variability also inherently captures the latitude in OAR dose trade-off extents permissible in the clinical environment. In this research, we utilize this inherent variability to our advantage to design a Machine Learning framework for head-and-neck treatment planning that allows dose distribution adjustability. We achieve this adjustability by proposing three models, namely \emph{“base”}, \emph{“lower-extreme”}, and \emph{“upper-extreme”} models. Here, rather than generating the typical single average \emph{“base”} Machine Learning solution ([6, 7]), we generate multiple paired solutions, with each pair representing a “lower-extreme” and “upper-extreme” solution for a specific OAR. The lower- and upper-extreme solutions are dose distributions in the entire 3D domain that incidentally also have the lowest and highest doses for the specific OAR, mimicking the clinically acceptable range seen in the training set. In particular, first, the \emph{“base model”} is obtained through training over all data points. For every OAR, we categorize each training dataset case into an upper- or lower-extreme dataset based on the training data planned dose being above or below, respectively, a threshold of the dose proposed by the base model for the OAR. Then, \emph{“lower-extreme”} and \emph{“upper-extreme”} models for the OAR are trained over lower- and upper-extreme datasets, respectively. Given the lower-extreme and upper-extreme model proposed dose distributions, planners and clinicians can combine the dose distributions into a single dose distribution by adjusting weighting parameters in real-time to allow lower or higher dose to selected OARs.

\section{Methods and Materials}\label{sec:methods}
In this section, we provide details about the dataset, feature engineering, and the Machine Learning model we utilize.
\subsection{Dataset}
Our dataset is comprised of the treatment planning data for 130 patients with head-and-neck cancer, previously treated with radiation therapy at the North Carolina Cancer Hospital, University of North Carolina, Chapel Hill.  The following data on each patient are included: the contours of OARs/targets (drawn on every slice of the patient’s CT set) and the clinically planned dose distribution (CDD) that encapsulates both the optimization algorithm and human expert planner manipulations. We consider brainstem, spinal cord, larynx, left parotid, and right parotid as the OARs. The targets, designated as Planning Target Volumes (PTV),  are High Risk PTV (HR-PTV) and/or Intermediate Risk PTV (IR-PTV) and/or Standard Risk PTV (SR-PTV).  The SR-PTV receives the lowest prescription dose, the IR-PTV (contained within the SR-PTV) receives a higher prescription dose, and the HR-PTV (contained within the IR-PTV) receives the highest prescription dose. As such, the doses to the PTVs reflects the level of suspicion of the presence of cancer. The number of PTVs varies based on patient and ranges from just one PTV (High Risk PTV) to all three PTVs.  The dataset was randomly divided into 100 training patients and 30 testing patients. Tables \eqref{tab:sites} and \eqref{tab:prs} list the treatment sites and prescriptions for the training and test sets, respectively.  In Table \eqref{tab:prs}, the number of prescription doses in each row of the left column corresponds to the number of targets. 
\begin{table}[t!]
\begin{minipage}{.48\linewidth}
	\caption{head-and-neck   treatment sites for the training and testing set cases}
	{\small
	\begin{tabular}{lcc}
		\toprule
		Treatment Site                       & \# Training          & \# Testing      \\\midrule
		Neck                                 & 3                              & 2                         \\
		Tonsil                               & 11                             & 5                         \\
		Tonsil and Neck                      & 3                              & 3                         \\
		Base of Tongue                       & 10                             & 4                         \\
		Base of Tongue and Neck              & 7                              & 1                         \\
		Larynx                               & 4                              & 1                         \\
		Larynx and Neck                      & 3                              & 3                         \\
		Parotid                              & 7                              & 1                         \\
		Parotid and Neck                     & 2                              &                           \\
		Oral Cavity                          & 3                              & 3                         \\
		Oral Cavity and Neck                 & 5                              & 3                         \\
		Soft Palate                          & 1                              &                           \\
		Floor of Mouth                       & 1                              &                           \\
		Floor of Mouth and Neck              & 3                              &                           \\
		Floor of mouth and Mandible          & 1                              &                           \\
		Mandible and Neck                    &                                & 1                         \\
		Tongue                               & 5                              &                           \\
		Tongue and Neck                      & 2                              & 1                         \\
		Hard Palate                          & 1                              &                           \\
		Soft Palate                          & 2                              &                           \\
		Subglottis                           & 1                              &                           \\
		Supraglottis                         & 1                              & 1                         \\
		Oropharynx                           & 4                              &                           \\
		Oropharynx and Neck                  & 3                              &                           \\
		Nasopharynx                          &                                & 1                         \\
		Nasopharynx and Neck                 & 4                              &                           \\
		Hypopharynx                          & 1                              &                           \\
		Hypopharynx and Neck                 & 2                              &                           \\
		Sinonasal                            & 4                              &                           \\
		Sinonasal and Neck                   & 3                              &                           \\
		Cheek and Neck                       & 1                              &                           \\
		Thyroid and Neck                     & 1                              &                           \\
		Esophagus and Neck                   & 1                              &                          \\\bottomrule
	\end{tabular}
}
\label{tab:sites}
\end{minipage}
\hfill\hspace{2.5cm}
\begin{minipage}{0.48\linewidth}
	\caption{head-and-neck treatment prescriptions for training and   testing set cases}
	{\small
	\begin{tabular}{lll}
		\toprule
		Prescription (Gy)               & \# Training               & \# Testing              \\ \midrule
		48, 44                          & 1                                   &                                \\
		60                              & 2                                   & 1                              \\
		60, 50                          & 14                                  & 7                              \\
		60, 45.6                        & 2                                   & 1                              \\
		60, 52                          & 1                                   &                                \\
		60, 54                          & 17                                  & 7                              \\
		60, 54, 50                      &                                     & 1                              \\
		66                              & 4                                   &                                \\
		66, 50                          & 4                                   &                                \\
		66, 54                          & 9                                   & 1                              \\
		66, 60,54                       & 10                                  & 4                              \\
		70, 50                          & 1                                   &                                \\
		70, 54                          & 30                                  & 5                              \\
		70, 60, 54                      &                                     & 1                              \\
		70, 63, 54                      & 1                                   &                                \\
		72, 54                          & 1                                   & 2                              \\
		74.4, 45.6                      & 3                                   &                                \\ \bottomrule
	\end{tabular}
}
\label{tab:prs}
\end{minipage}
\end{table}

\subsection{U-Net Model}
In this work,  we use a U-Net convolutional-based deep neural network architecture  \cite{ronneberger2015u}.   U-Net has been used extensively and has provided promising results in a wide range of medical applications such as image segmentation, outcome prediction, dose quantification, and radiation adaptation  \cite{nguyen2019feasibility,sadeghnejad2020fast}.

\begin{figure}[t!]
	\centering
	\begin{tabular}{c}
		\includegraphics[trim={0 2.3cm 0 0},width=15cm]{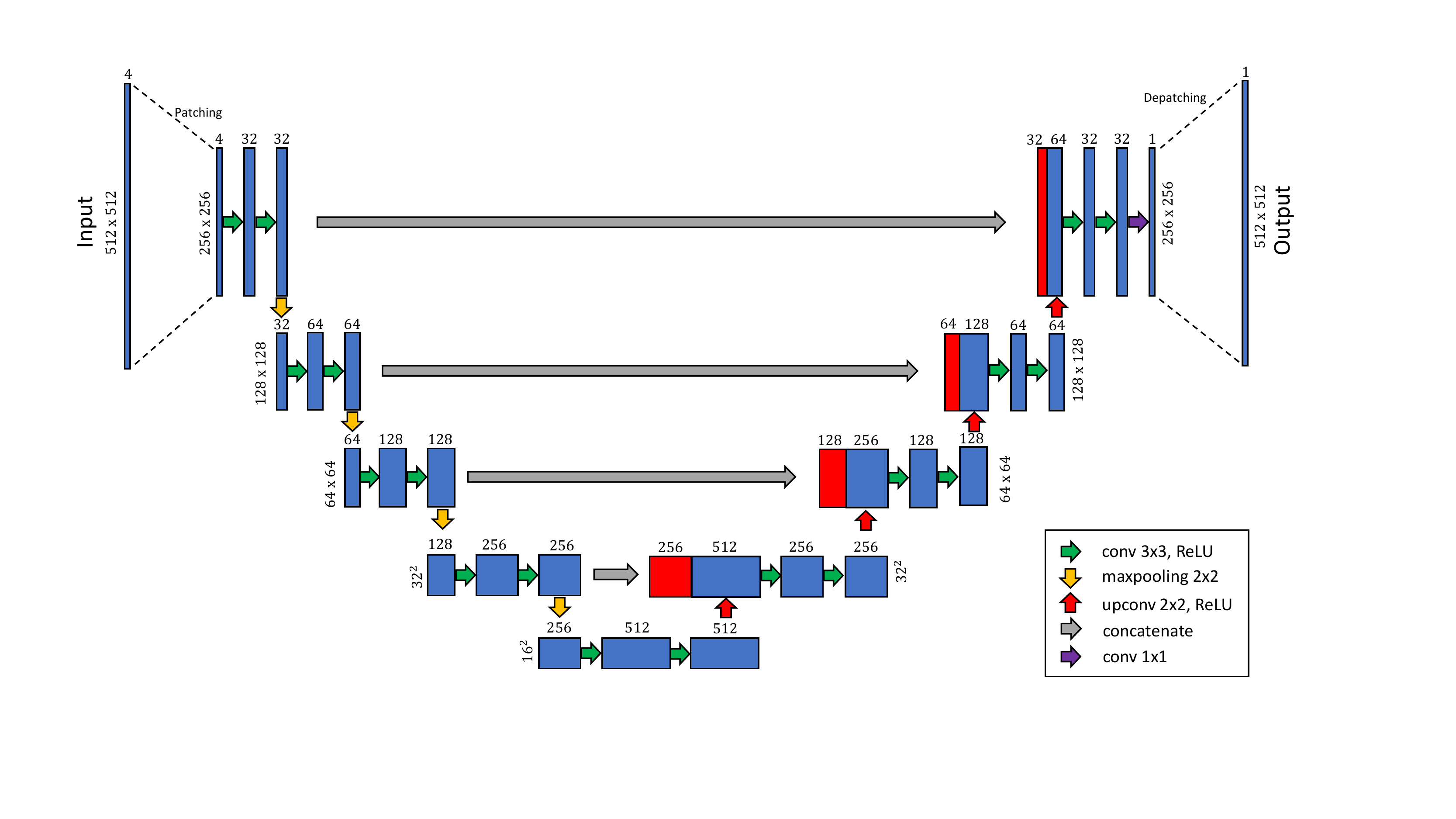}
	\end{tabular}
	\caption{U-Net:  The left side of this U-shape model is the contracting part, which consists of two copies of two convolutional layers with kernel size $3\times 3$, and one max-pooling layer with kernel size $2\times 2$.  The bottom part is called bottleneck and uses two $3\times 3$ CNN layers followed by $2\times 2$ up-convolution layer. The expanding part, which is the right side, consists of repeated blocks each containing two convolutional layers followed by an up-convolutional layer.}
	\label{fig:model}
\end{figure}

 The U-Net model includes a sequence of convolution, max-pooling, and transposed convolution operations in order to extract local and global features from the input and to return the predicted dose distribution (Fig.\eqref{fig:model}). In particular, the U-Net consists of two main blocks, namely the contracting block and the expanding block. The left side of this U-shape model is the contracting block, which consists of two copies of two convolutional layers with kernel size $3\time 3$, and one max-pooling layer with kernel size $2\times 2$. The activation function for convolutional layers is set to ReLU. The expanding part, which is the right side of the U-shape, consists of repeated blocks, each containing two convolutional layers followed by an up-convolutional layer.

\subsection{Data and Feature Engineering} 
The 3D dose distribution for every patient consists of a concatenation of 2D slices of size $512\times 512$. Each point in this 3D tensor contains dose values in Gray (Gy) provided by the CDD. Further, for every patient, the prescribed dose values for targets and the limit dose values (average or max) for OARs   are available. Based on this information, we build the input feature matrix which is a four-channel image stack as follows: (i) prescribed dose values for target, (ii) type of tissue, (iii) dose limit value for OARs, (iv) distance to  the closest target set for each pixel. Regarding the type of tissue, encoded in the second channel, we consider five organs at risk (OARs), namely brainstem, spinal cord, larynx, left parotid, and right parotid, and three types of targets, namely SR-PTV, IR-PTV, and HR-PTV. Stacking these four channels on top of each other, we have a $512\times 512\times 4$ tensor for each 2D slice. For the labels, we simply consider the CDD from the treatment plan, which is specifically a matrix of size $512\times 512$ for each 2D slice.  Further, we divide the slices of size $512\times 512$ into smaller patches with a predetermined stride. Patching big images into smaller pieces has two benefits: (i) it makes the input size smaller and more tractable, and (ii) it helps to provide more training data.

\subsection{Base, Lower-extreme and Upper-extreme Models }\label{sec:adaptive_model}

\begin{figure}
	\centering
	\begin{tabular}{lll}
		\includegraphics[width=5cm]{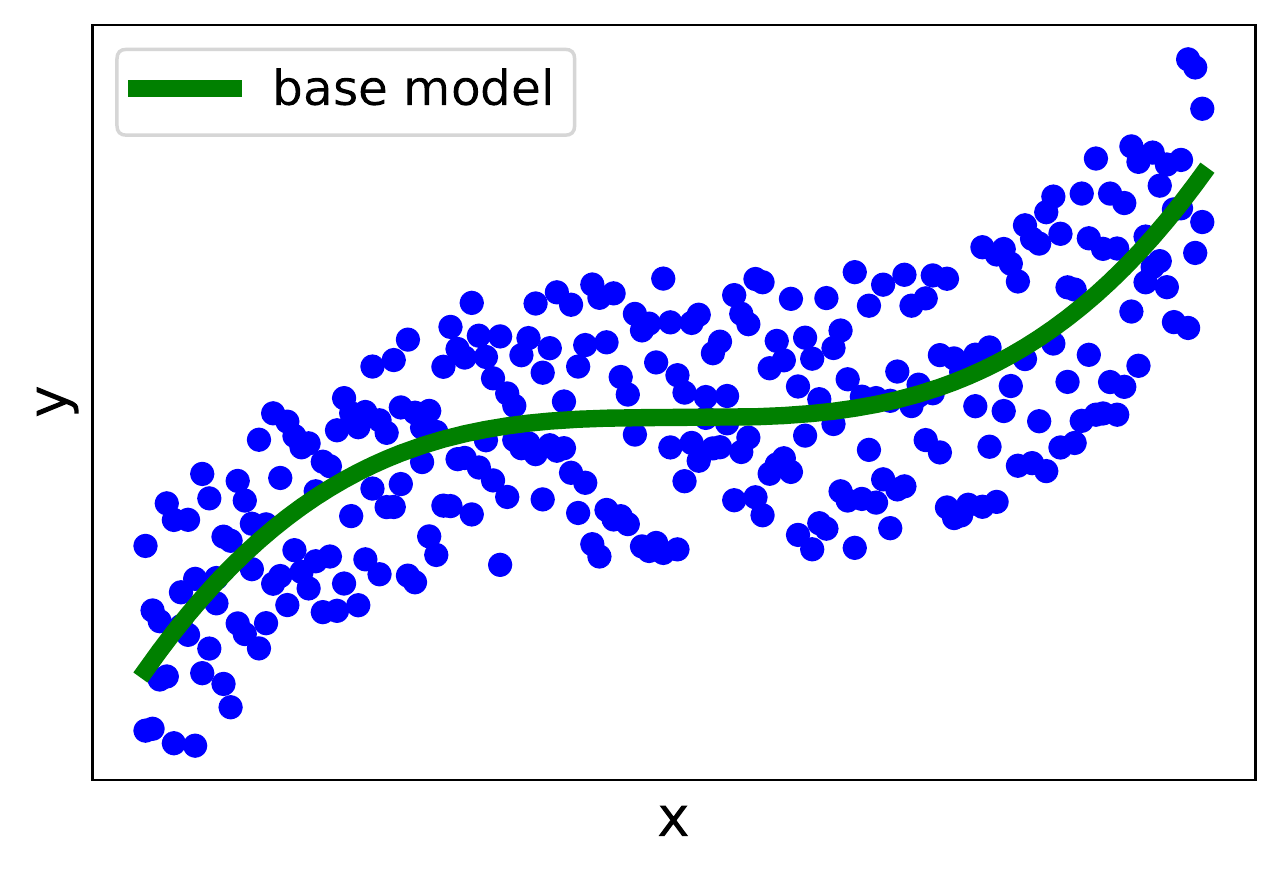}&
		\includegraphics[width=5cm]{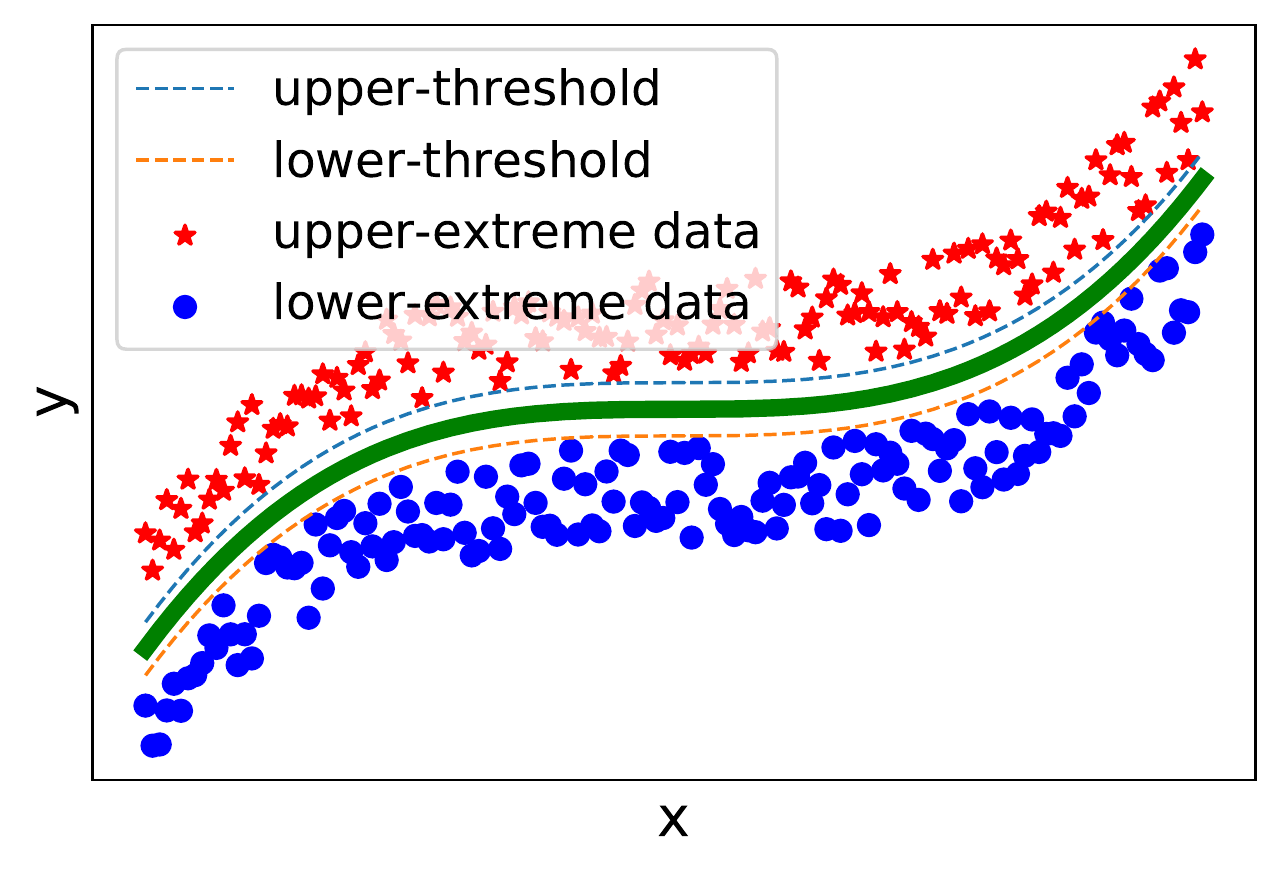}&
		\includegraphics[width=5cm]{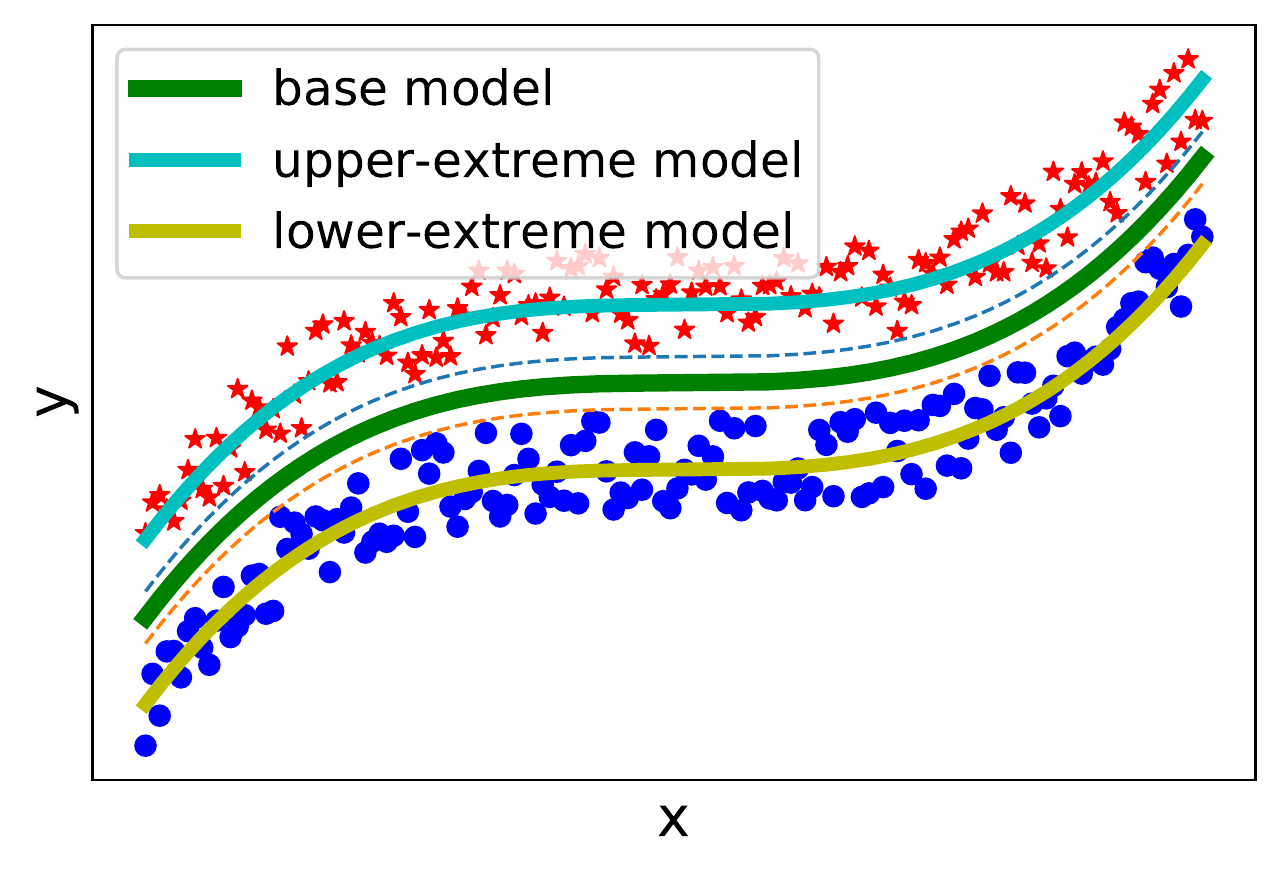}
	\end{tabular}
	\caption{(a) Schematic representation of the base model’s dose prediction for a particular OAR (green line) superimposed on the CDD (clinically planned) doses for the training set (blue points);  (b) red and blue points  represent the upper- and lower-extreme training cases that are used to train the upper- and lower-extreme models, respectively; Dashed lines represent the border line for $70\%$ above and below the base model.  (c) the upper- and lower-extreme model predictions are shown as blue and light-green lines, respectively. }
	\label{fig:base_model}
\end{figure}

In this subsection, we provide more details on the development of “base”, “lower-extreme”  and “upper-extreme” models.  First, a base U-Net model is created by training on all patients in the training set. This base model is analogous to the single model that is conventionally seen in the literature to predict the dose distribution \cite{nguyen2019generating, bohara2020using}. Next, to create the lower-extreme and upper-extreme models, the base model is utilized to predict the dose distribution for all patients in the training set. 
For each OAR, the training dataset is categorized into lower- and upper-extreme subsets by comparing the base model predicted dose values to the CDD dose values. In figure \ref{fig:base_model} (a), left panel, we schematically illustrate, for a particular OAR, the predicted doses (green line) using the base model, and the actual CDD OAR doses (blue points). CDD with dose values above the predicted dose line represents those patients with upper-extreme doses (red points in figure \ref{fig:base_model} (b)), while CDD dose points below the green line represent patients with lower-extreme doses (blue points in in figure \ref{fig:base_model} (b)).  The lower-extreme points represent cases where the planner pushed for greater OAR sparing compared to the average prediction from the base model. Conversely, the upper-extreme points represent cases where the planner attributed lower importance to the OAR in order to obtain greater sparing elsewhere. The U-Net model is next trained on the upper-extreme and lower extreme cases to generate the upper-extreme and lower-extreme models, respectively, for the particular OAR; see figure \ref{fig:base_model}(c). To provide a greater range between the upper- and lower-extreme models, only cases that are $30\%$ above/below the base model prediction are used to train the upper-/lower-extreme models (see dashed border lines in figures \ref{fig:base_model}(b), \ref{fig:base_model}(c)). The upper- and lower-extreme model pairs are intended to produce dose distributions with higher and lower doses, respectively, to a particular OAR.   This process is repeated for each OAR, yielding an upper-extreme and lower-extreme model pair for each OAR.  The upper- and lower-extreme models automatically capture the dose trade-off between OARs, since these trade-offs are inherent to the training set of patient plans in the upper-extreme and lower-extreme training cases.

Since the upper- and lower-extreme sets for each OAR contain fewer patients’ data compared to the original base model data, one can use  {\it transfer learning} to initialize the extreme models with the base model. Transfer learning refers to the case where a trained model is exploited for training a different model in order to improve training efficiency and speed \cite[Chapter 11]{OlivasHandbook}.

The model dose distribution (MDD) prediction from an upper-extreme and lower-extreme model pair for a particular OAR ($OAR_i$) can be combined into a single prediction that spans the dose range
\begin{align}\label{eq:mdd}
\text{MDD}^{{OAR}_{i}} = \alpha_i \times\text{MDD}^{{OAR}_{i}}_{LE} + (1-\alpha_i )\times\text{MDD}^{{OAR}_{i}}_{UE},
\end{align}
where $\text{MDD}^{{OAR}_{i}}_{LE}$ and $\text{MDD}^{{OAR}_{i}}_{UE}$  are the lower- and upper-extreme model dose distribution predictions in the entire 3D domain, and $0\leq\alpha_i\leq 1$ is the “risk parameter” associated with the $OAR_i$. For a particular patient, if $OAR_i$ is deemed to be high risk, a larger value of $\alpha_i$ may be chosen to increase the influence of the lower-extreme model. Note that the linear interpolation between two optimal solutions is not necessarily an optimal solution.  However, the range between the upper-extreme and lower-extreme model solutions is modest because the dose trade-off between OARs is restricted to a clinically acceptable range. Consequently, we can reasonably assume that the linear interpolation between upper-extreme and lower-extreme model solutions is close to optimal.

The dose distributions corresponding to different OARs are then combined to produce a single dose distribution:
\begin{align}\label{eq:final_plan}
\text{MDD} = \sum_{i}w_i \text{MDD}^{{OAR}_{i}}/\sum_i w_i,
\end{align}
where $w_i$'s are ``organ weights'' associated with the importance of $OAR_i$ compared to the other OARs. Thus, parameter $\alpha_i$ and $w_i$ can be adjusted in real-time to tailor the dose distribution as best as possible.
\section{Results}\label{sec:results}


In this section, we provide details and results of numerical experimentation to demonstrate the proposed method. In the first part, we discuss training the base model using U-Net, and in the second part, we explore the extreme models for providing adjustable dose distributions. 

\subsection{Base Model}
As mentioned earlier, the original slices of size $512\times 512$  are divided into smaller patches. From experimentation, we selected a patch size of $128$ with stride $64$. To avoid over-fitting, a dropout regularization was added to the U-Net model with rate $0.1$ after each convolution layer. The Adam optimization algorithm with a learning rate of  $0.001$, and batch size $32$ was used to train the model. The training was run on  a server with two Nvidia Tesla V100 32 GB GPUs for 30 hours. 
We define the absolute and relative target/OAR dose error in Gy as (dose refers to either the maximum or average dose based on the target/OAR):


\begin{align}
\texttt{Abs-Error}&=|\texttt{dose~from~CDD - dose~predicted~by~U-Net}|\label{eq:abs_err}\\
\texttt{Rel-Error}&=\left|\frac{\texttt{dose~from~CDD - dose~predicted~by~U-Net}}{\texttt{dose~from~CDD}}\right|\times 100 \label{eq:rel_err}
\end{align}

\begin{figure}[htbp]
	\centering
	\begin{tabular}{cc}
	\includegraphics[width=50mm]{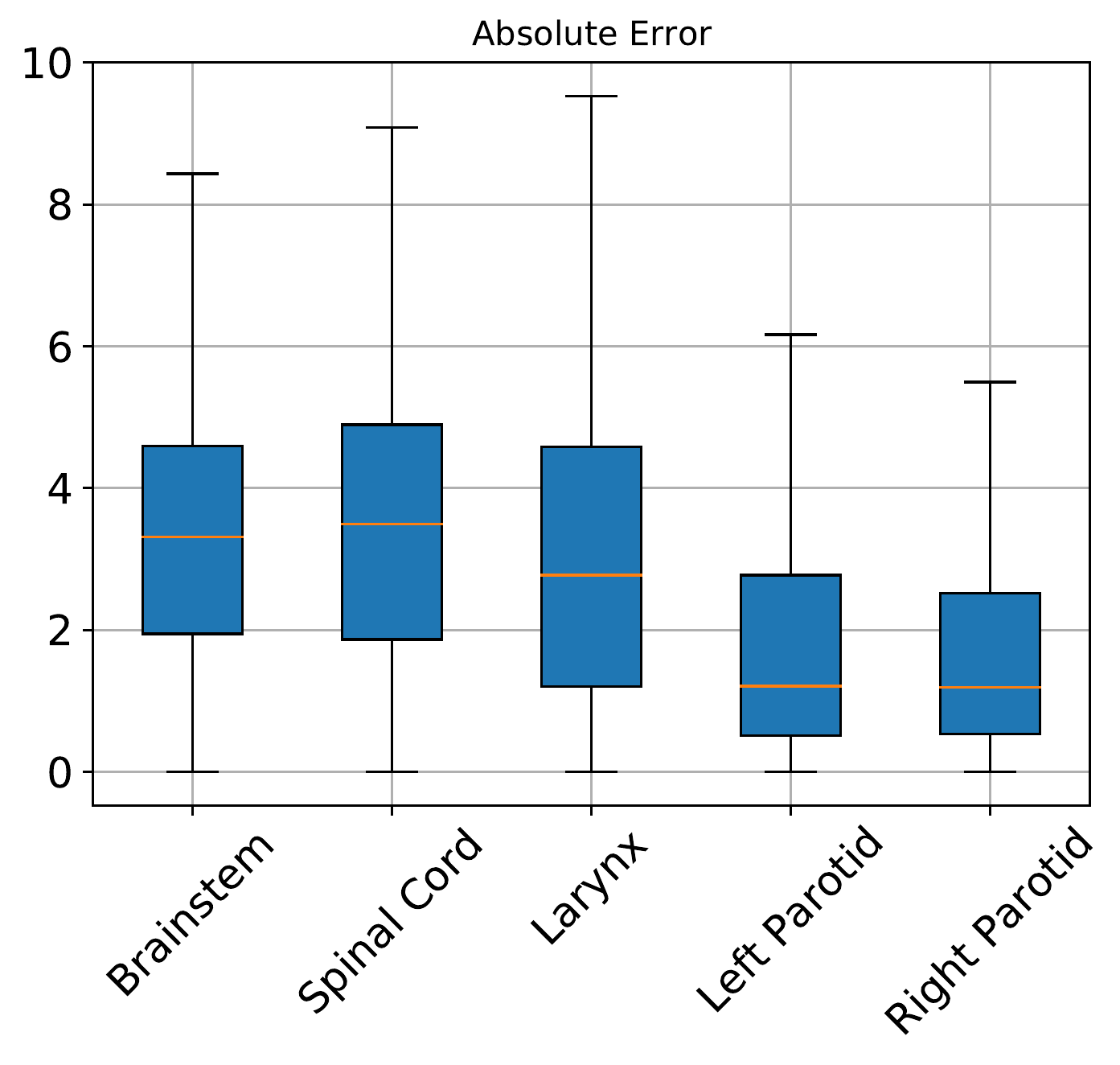} &\includegraphics[width=50mm]{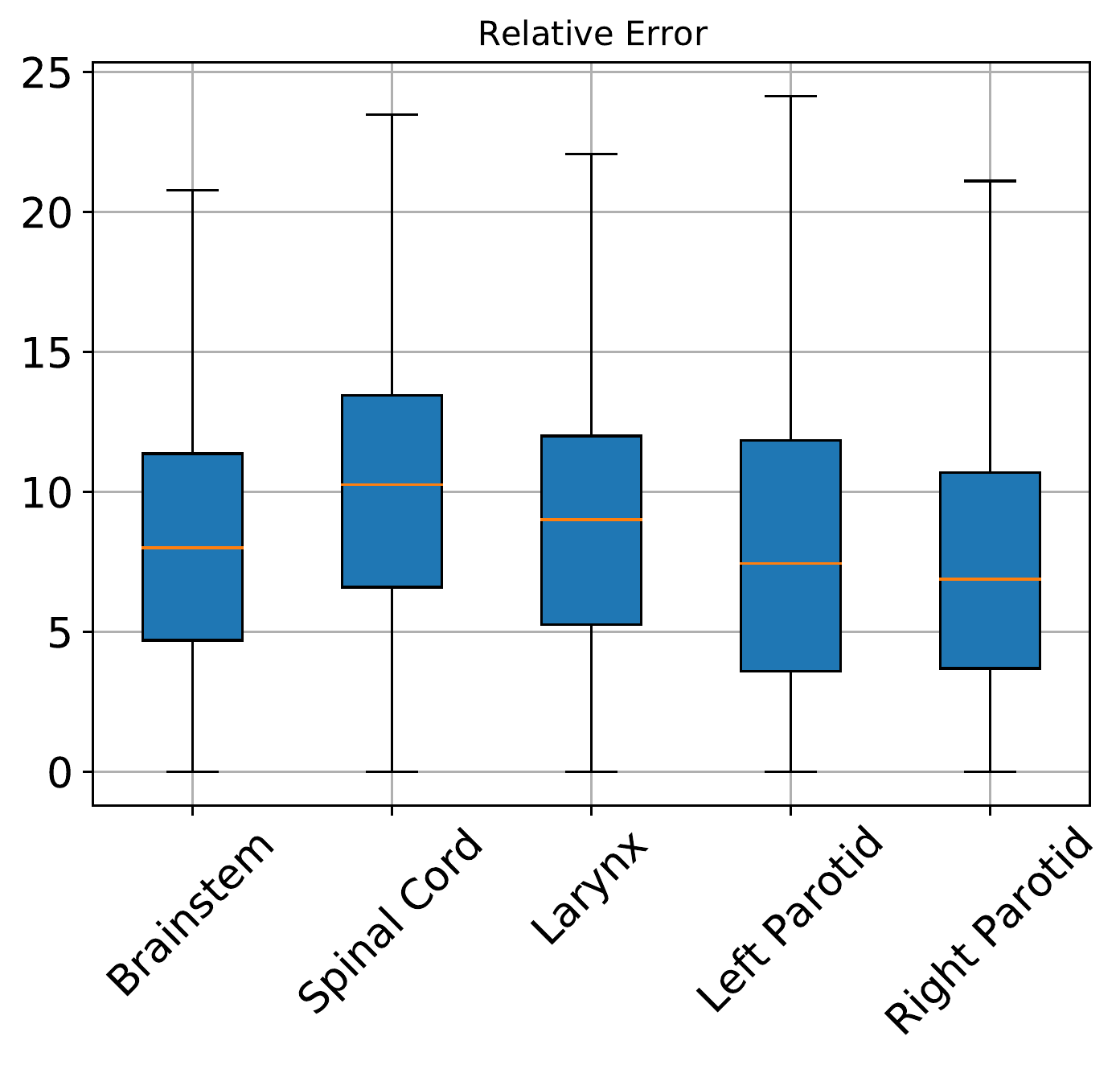}
	\end{tabular}
	\caption{\small Box plot of the absolute and relative errors for a randomly picked testing patient}
	\label{fig:box_plot}
\end{figure}

The mean and standard deviation values for the absolute and relative error are shown in Table \ref{tab:rel_err} for the test cases. The reported metrics for the targets and OARs are: average dose (PTV, larynx, left parotid and right parotid) and maximum dose (brainstem and spinal cord).  To reduce the uncertainties associated with reporting the maximum dose as the dose to a single pixel in an OAR, we instead report the maximum dose as the lowest dose to the highest $1\%$ of OAR pixels.  Compared to the CDD values, U-Net has reasonably small errors in predicting the dose values to targets— an average of $2.28\pm 1.48$ Gy and $7.92\pm 3.12 \%$ for absolute and relative error, respectively. For the OARs, U-Net errors are reasonably small for larynx and parotids.  However, they are a little higher for spinal cord and brainstem, owing to their relatively smaller sizes and the maximum dose being associated with just a few pixels. The Box plot of OAR errors is plotted in Figure \ref{fig:box_plot} to illustrate more statistical details of the errors.


\begin{table}[h]
	\small
	\centering
	\begin{adjustbox}{width=1\textwidth}
		\begin{tabular}{lcccccc}
			\hline
			& PTV(mean              &  \multicolumn{1}{c}{brainstem(max} & \multicolumn{1}{c}{spinal cord(max} & \multicolumn{1}{c}{larynx(mean} & \multicolumn{1}{c}{left parotid(mean}& \multicolumn{1}{c}{right parotid(mean}\\ 
			   &  dose)&   dose)&   dose)&  dose)&  dose)&  dose) \\ \hline
			Abs-Error (Gy)   & 5.82 $\pm$ 0.69&   2.18 $\pm$ 1.08 &   2.47 $\pm$ 1.38 &  2.48 $\pm$ 1.58 &  1.98 $\pm$ 1.67 & 1.55 $\pm$ 1.04  \\
			Rel-Error (\%) & 9.40 $\pm$ 0.55  & 6.51 $\pm$ 2.63 &  6.72 $\pm$ 3.32  &  6.73 $\pm$ 3.10 &  7.85 $\pm$ 2.26 & 6.94 $\pm$ 3.43\\ \hline
		\end{tabular}
	\end{adjustbox}
	\caption{\small The average error for targets and OARs in test patients.}
	\label{tab:rel_err}
\end{table}

\begin{figure}[h!]
	\centering
	\begin{tabular}{lll}
		\includegraphics[width=3.5cm]{ 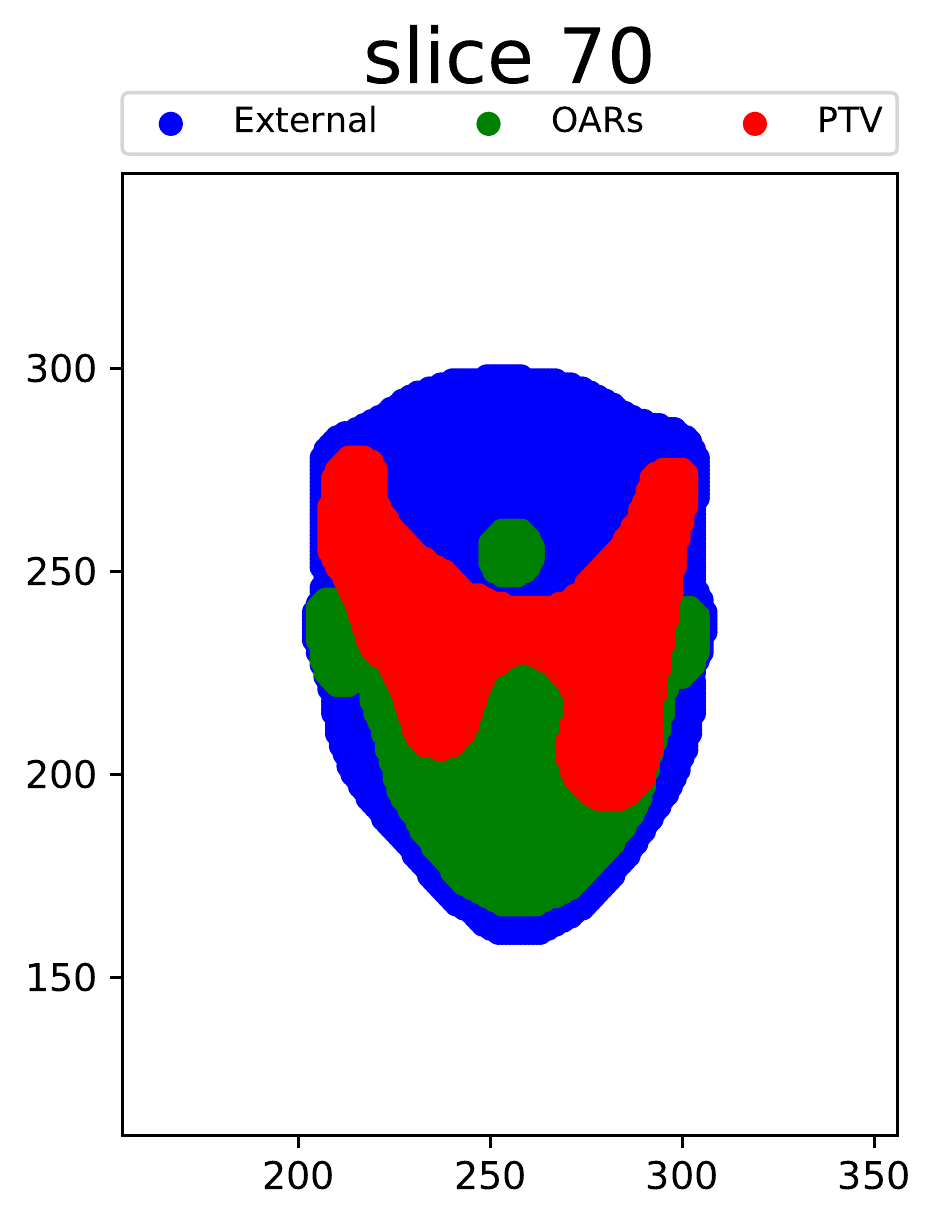}  & \includegraphics[width=3.5cm]{ 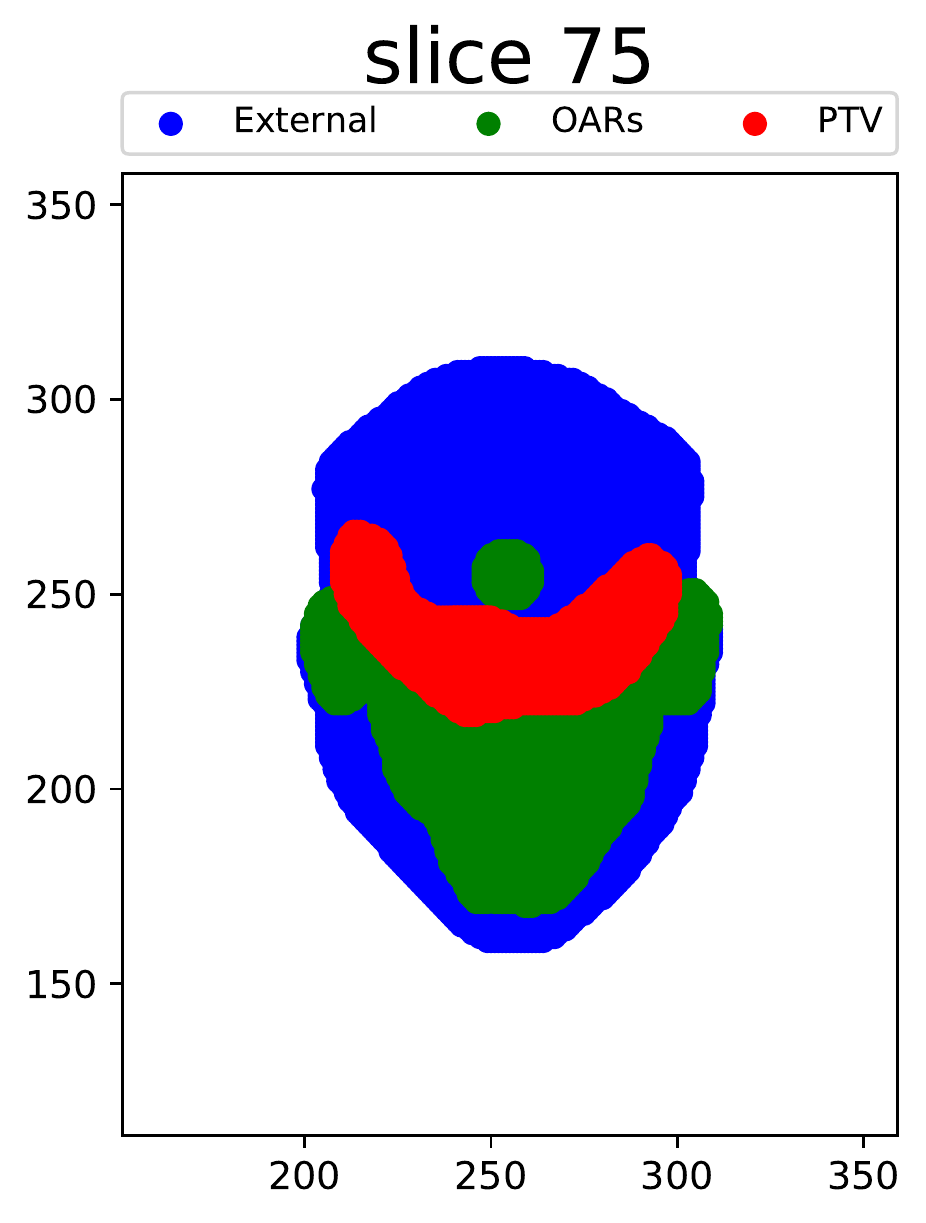}  & \includegraphics[width=3.5cm]{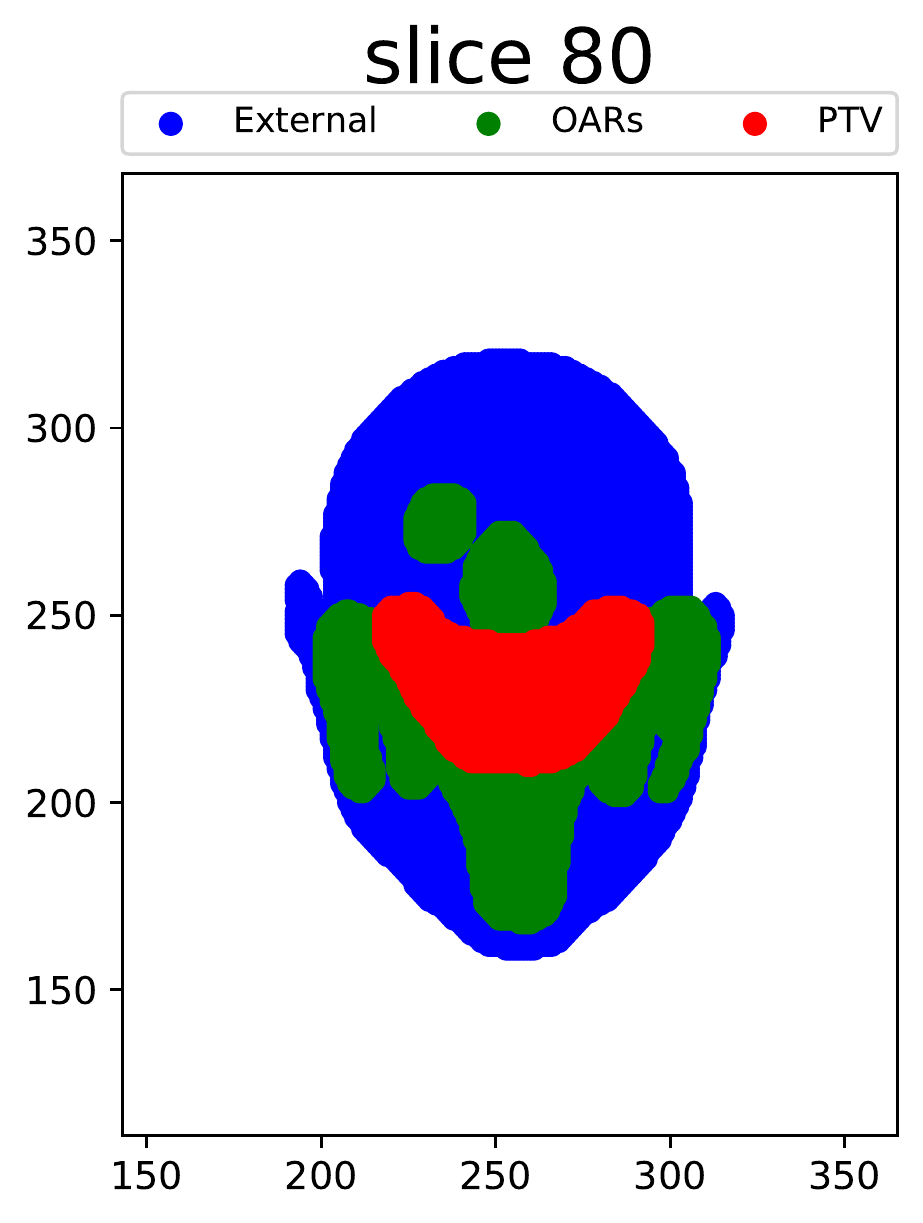} \\
		\includegraphics[width=4.cm, height=4.3cm]{ 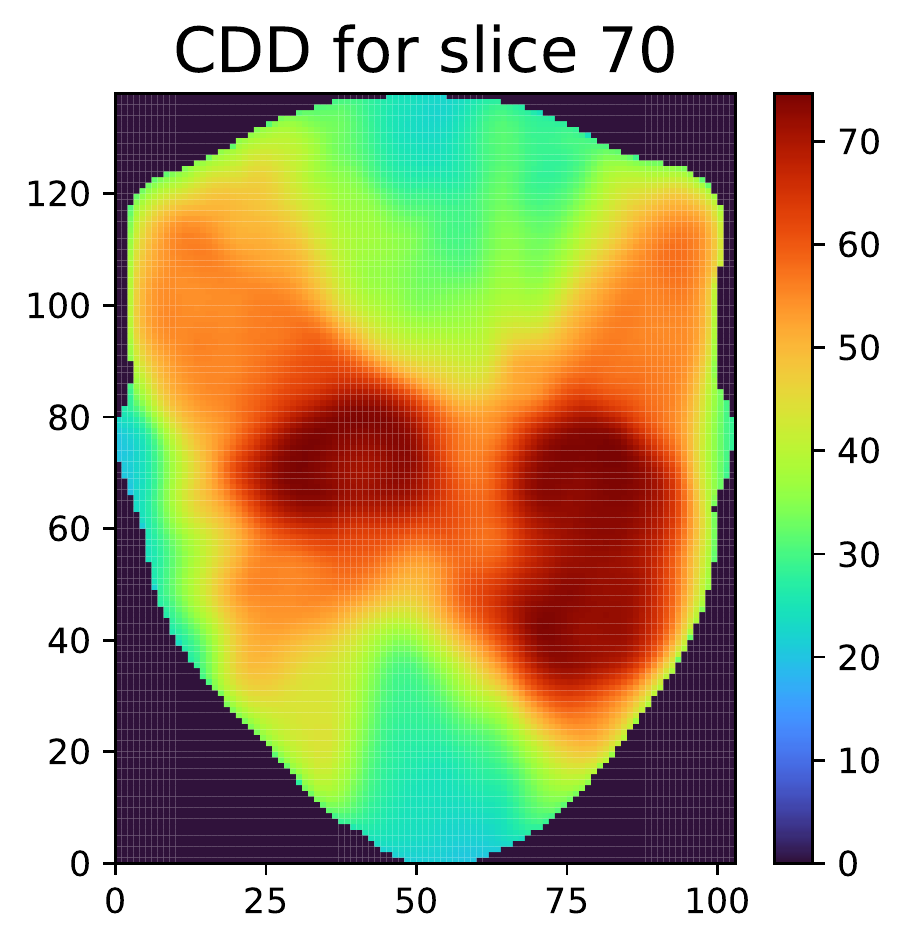} & \includegraphics[width=4cm, height=4.3cm]{ 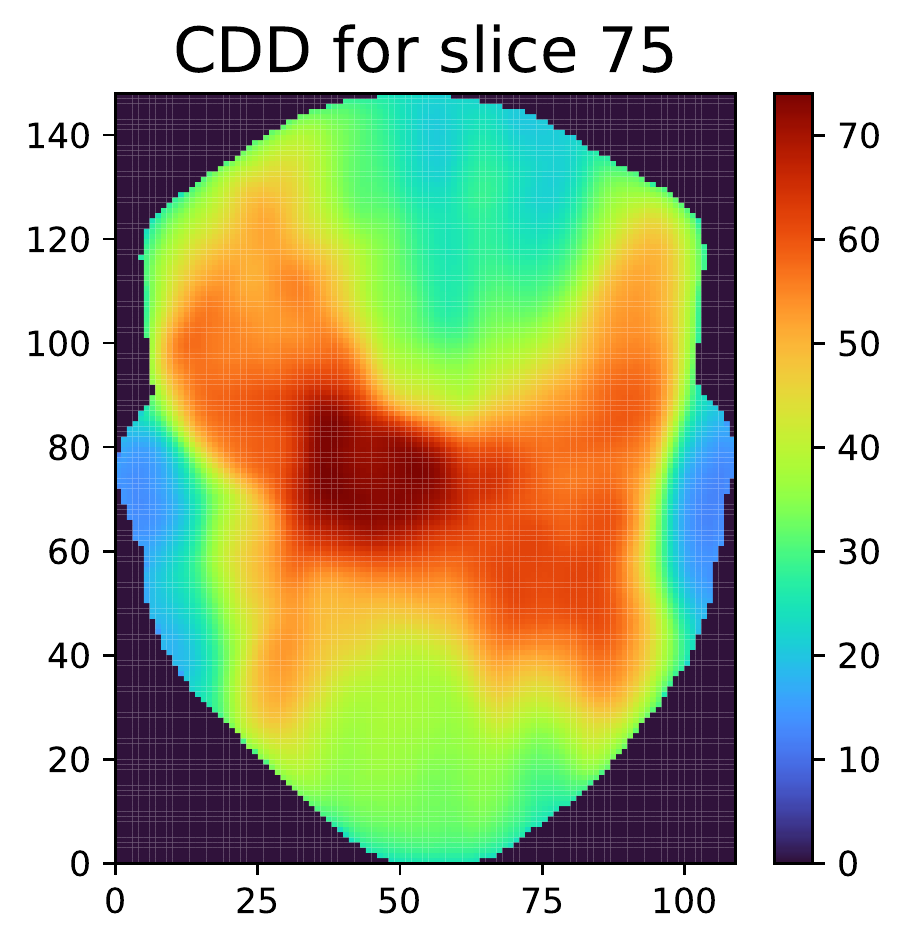}& \includegraphics[width=4.cm, height=4.3cm]{ 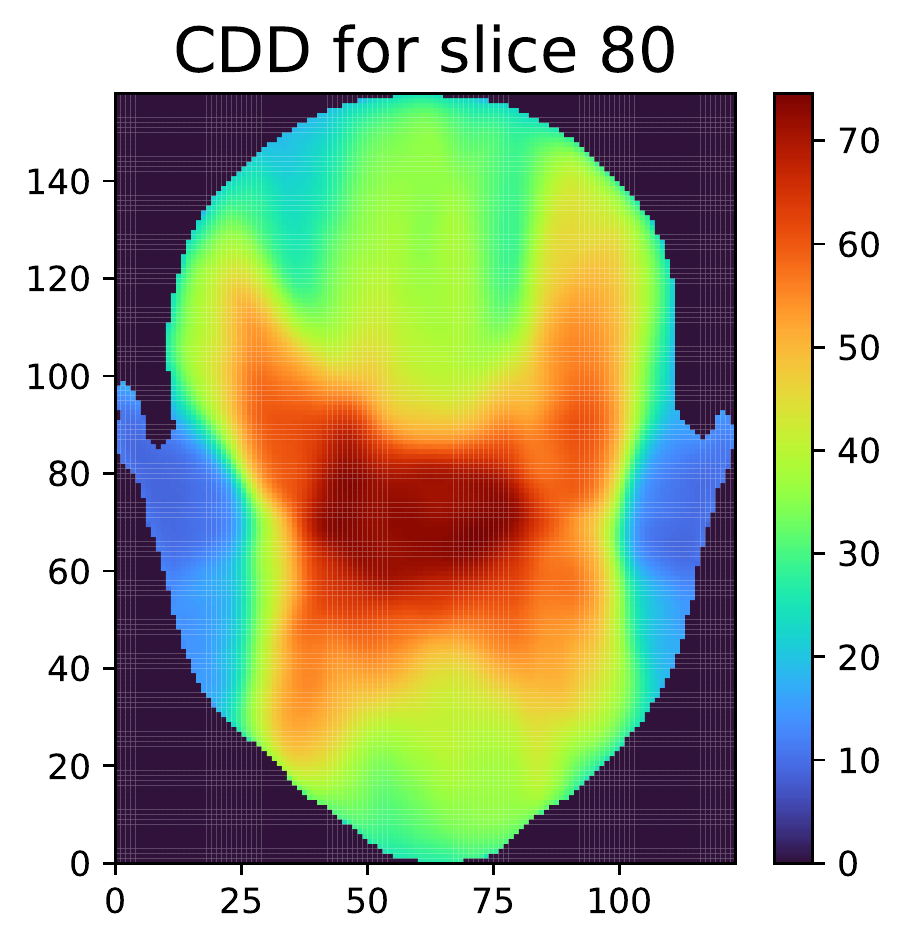}\\
		\includegraphics[width=4.cm, height=4.3cm]{ 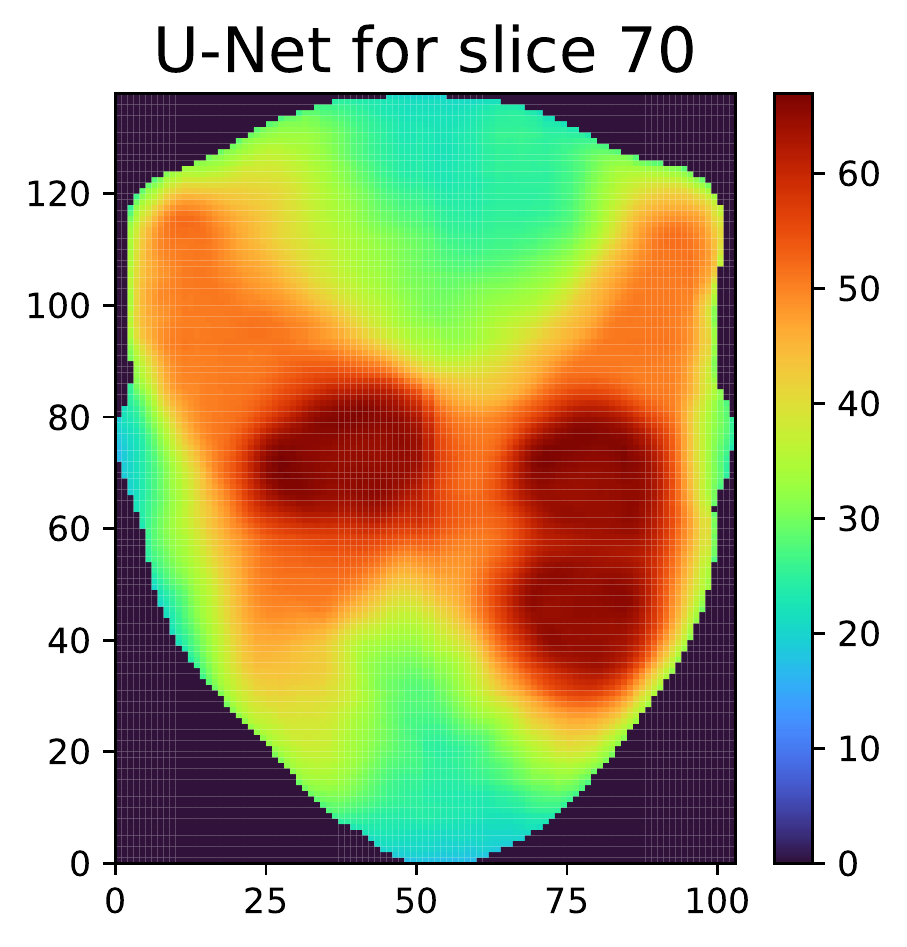} &\includegraphics[width=4.0cm, height=4.3cm]{ 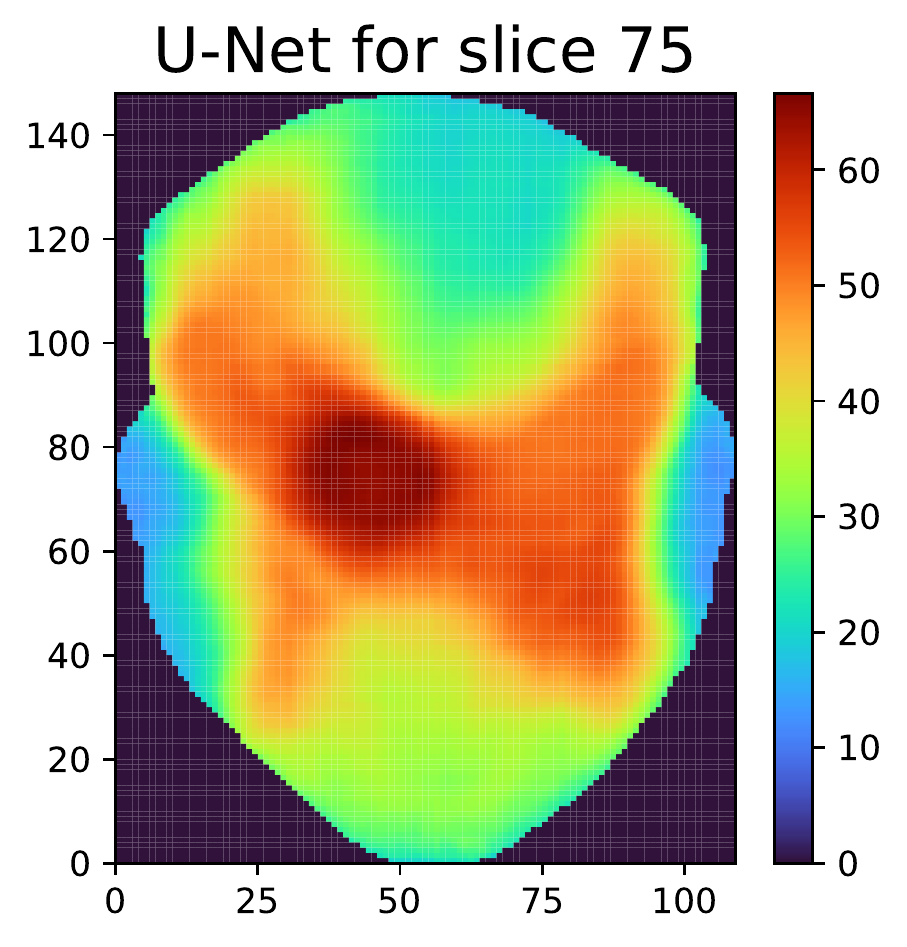} &\includegraphics[width=4.0cm, height=4.3cm]{ 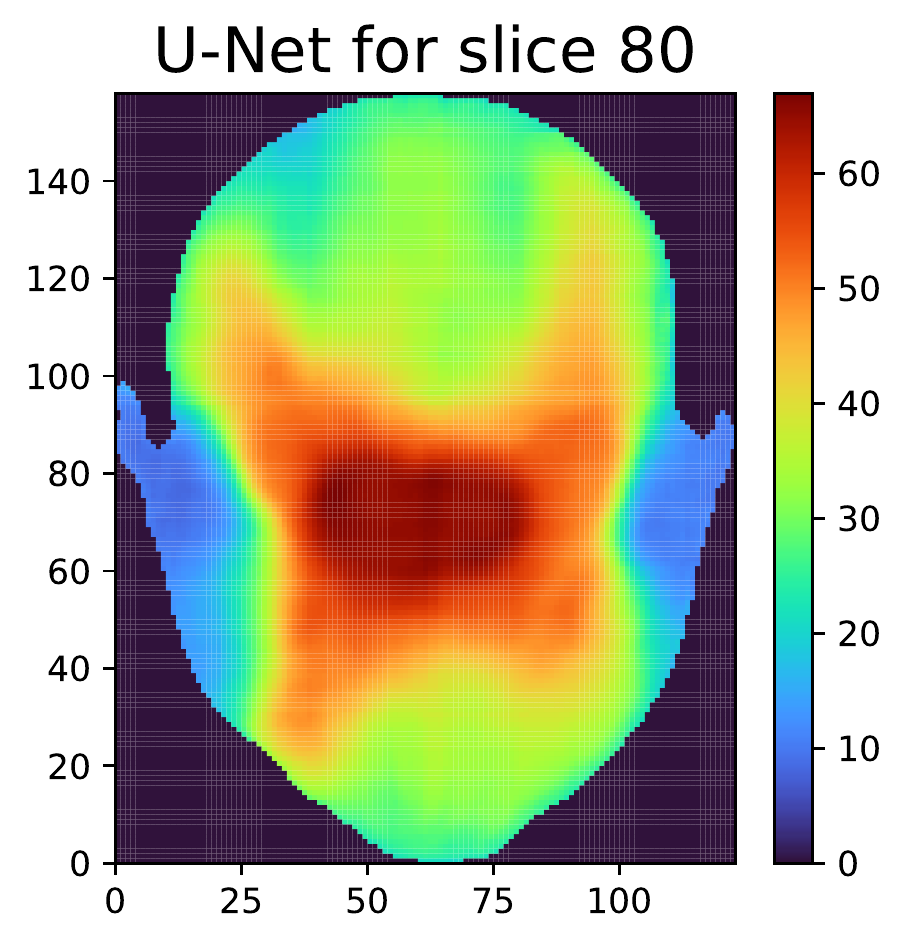}\\
	\end{tabular}
	\caption{\small Each column shows one specific slice. The first row is the contour representing different structures. The second row and the third row are the isodose lines of the dose distributions from CDD and U-Net model, respectively.}
	\label{fig:scatter}
\end{figure}

For visualization purposes, figure \ref{fig:scatter} shows the isodose lines of CDD and U-Net doses superimposed on the structure contours for a randomly selected test patient. In figure \ref{fig:scatter}, every column is associated with a specific slice. As illustrated, the U-Net model mimics the CDD reasonably well. For this test patient, figure \ref{fig:dvh_all_orgs} compares the CDD vs. U-Net cumulative dose-volume histogram (DVH) for targets and OARs (DVH plots the percentage volume (y-axis) of an organ receiving dose greater than or equal to the dose value on the x-axis). The comparison illustrates the ability of the U-Net model to mimic the CDD.

\subsection{Extreme Models}
\begin{figure}
\begin{minipage}{0.48\linewidth}
		\includegraphics[width=7.5cm]{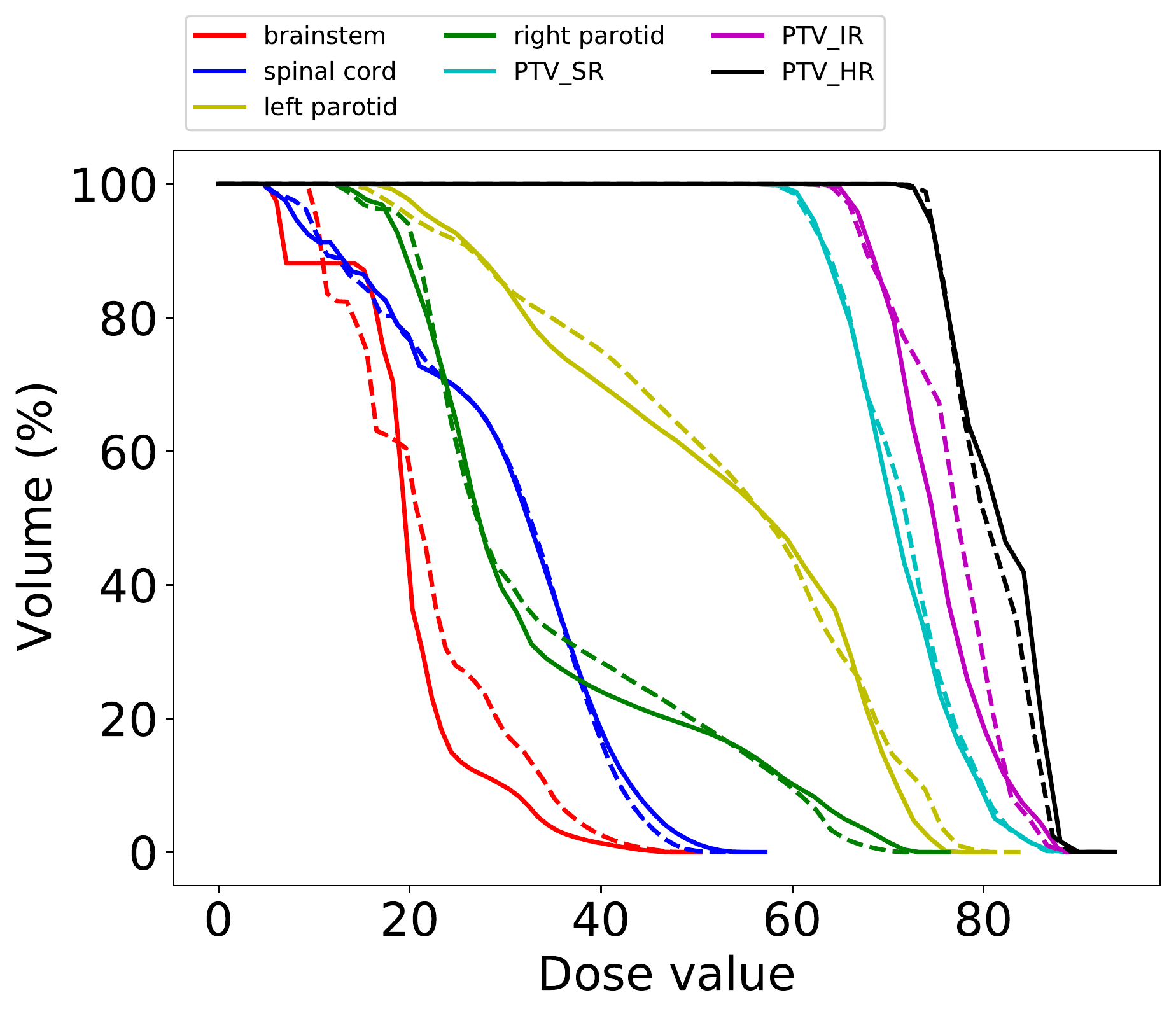}
		\caption{DVH plot CDD (dashed) vs. U-Net (solid).}
		\label{fig:dvh_all_orgs}
\end{minipage}
\hfill
\begin{minipage}{0.48\linewidth}
		\includegraphics[width=7.2cm]{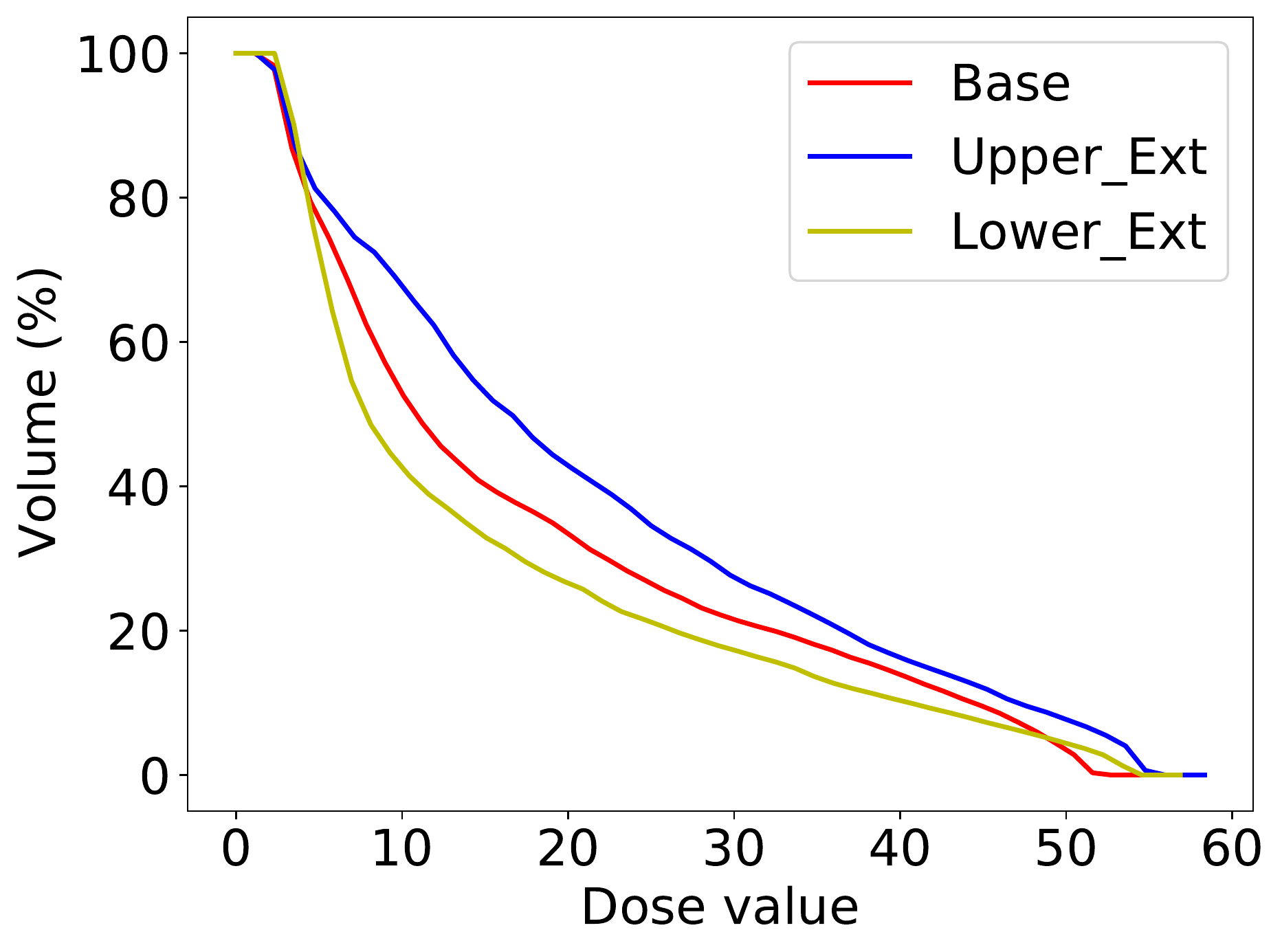}
		\caption{Right parotid DVH plots of base, upper- and lower-extreme models.}
		\label{fig:dvh_3_models}
\end{minipage}
\end{figure}

\begin{figure}[hh!]
	\centering
	\begin{tabular}{ll}
	\hspace{-.3cm}\includegraphics[width=7.5cm]{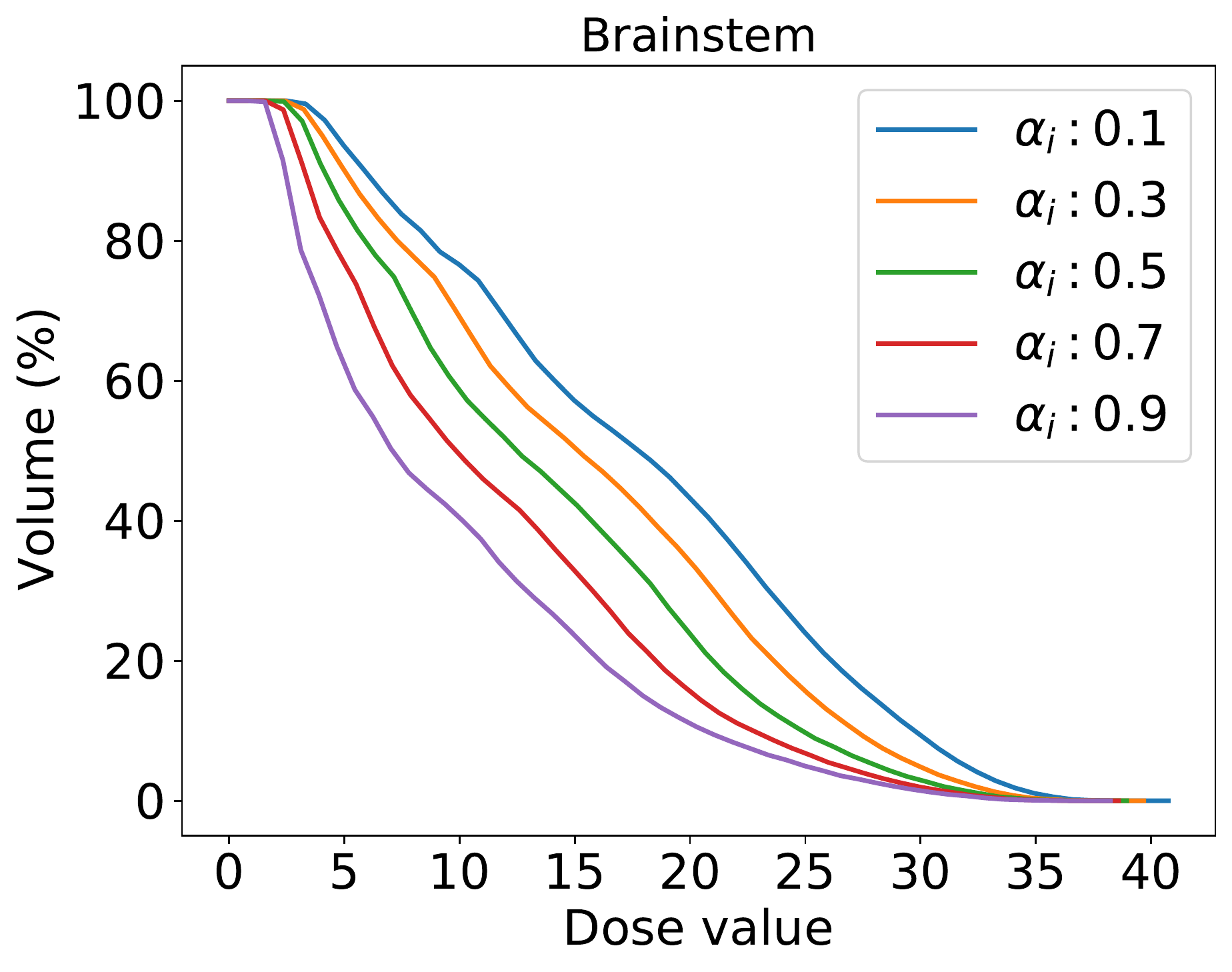}&	\includegraphics[width=7.5cm]{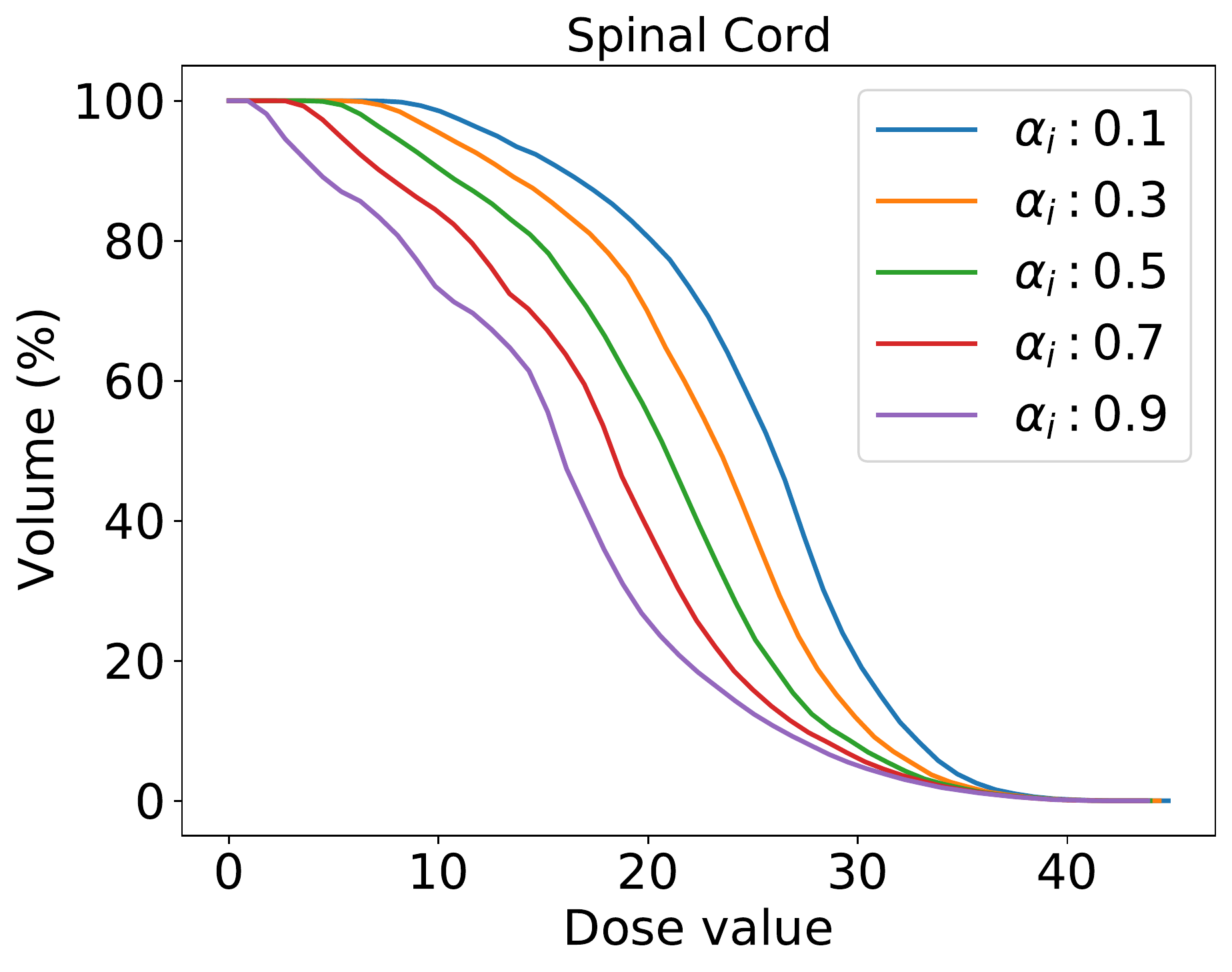}\\
	\hspace{-.3cm}\includegraphics[width=7.5cm]{ 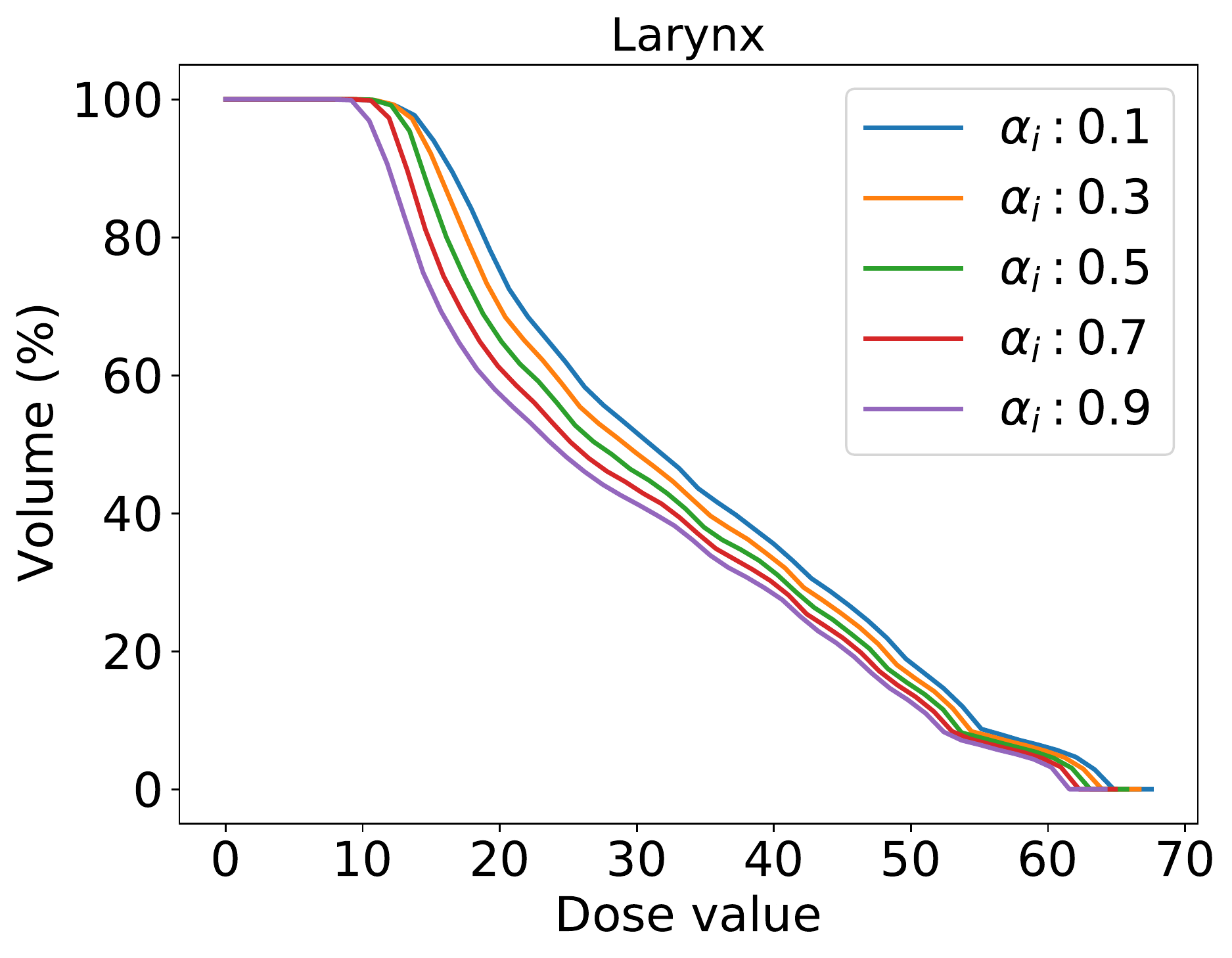}&
	\hspace{-.3cm}\includegraphics[width=7.5cm]{ 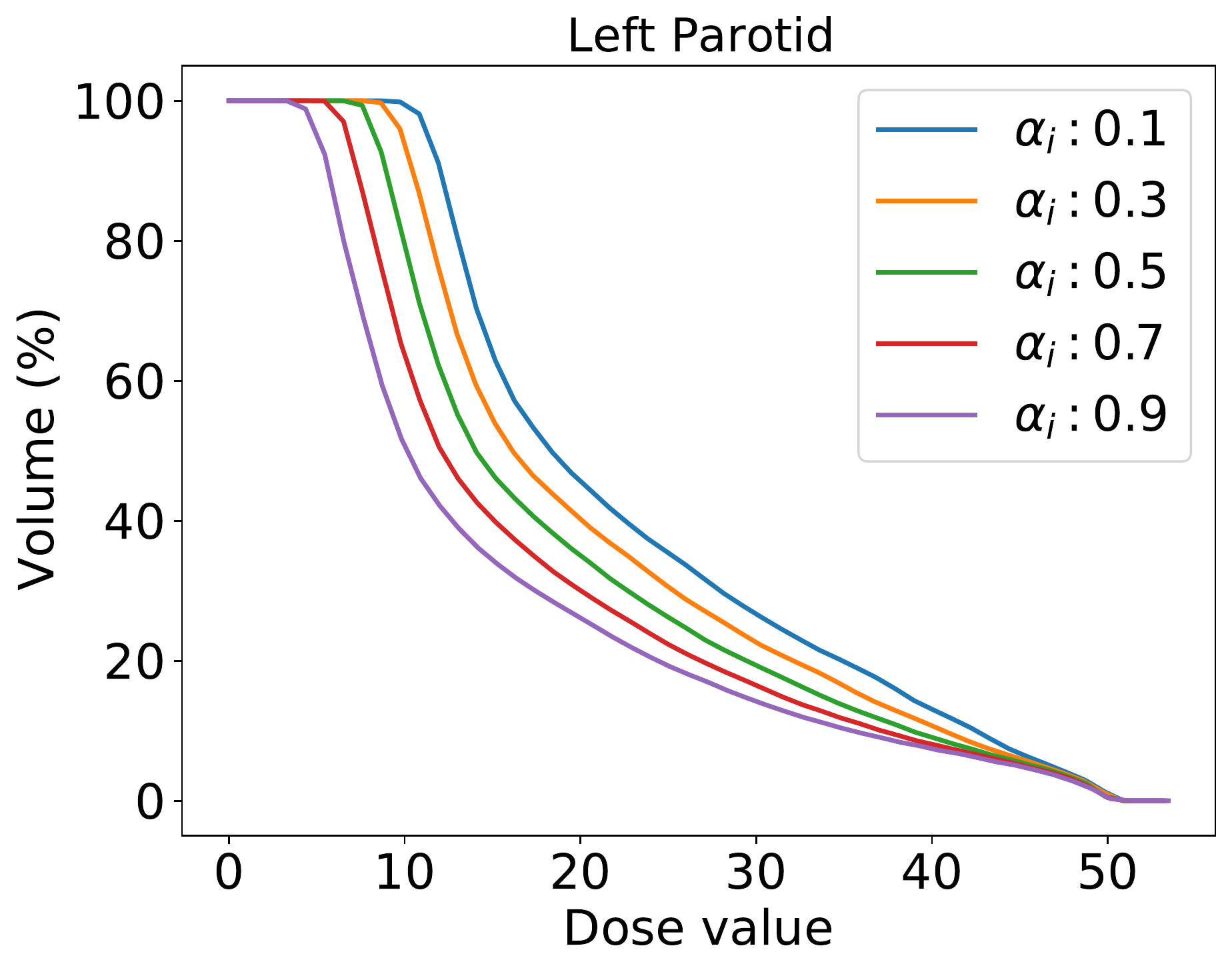}\\
	\includegraphics[width=7.5cm]{ 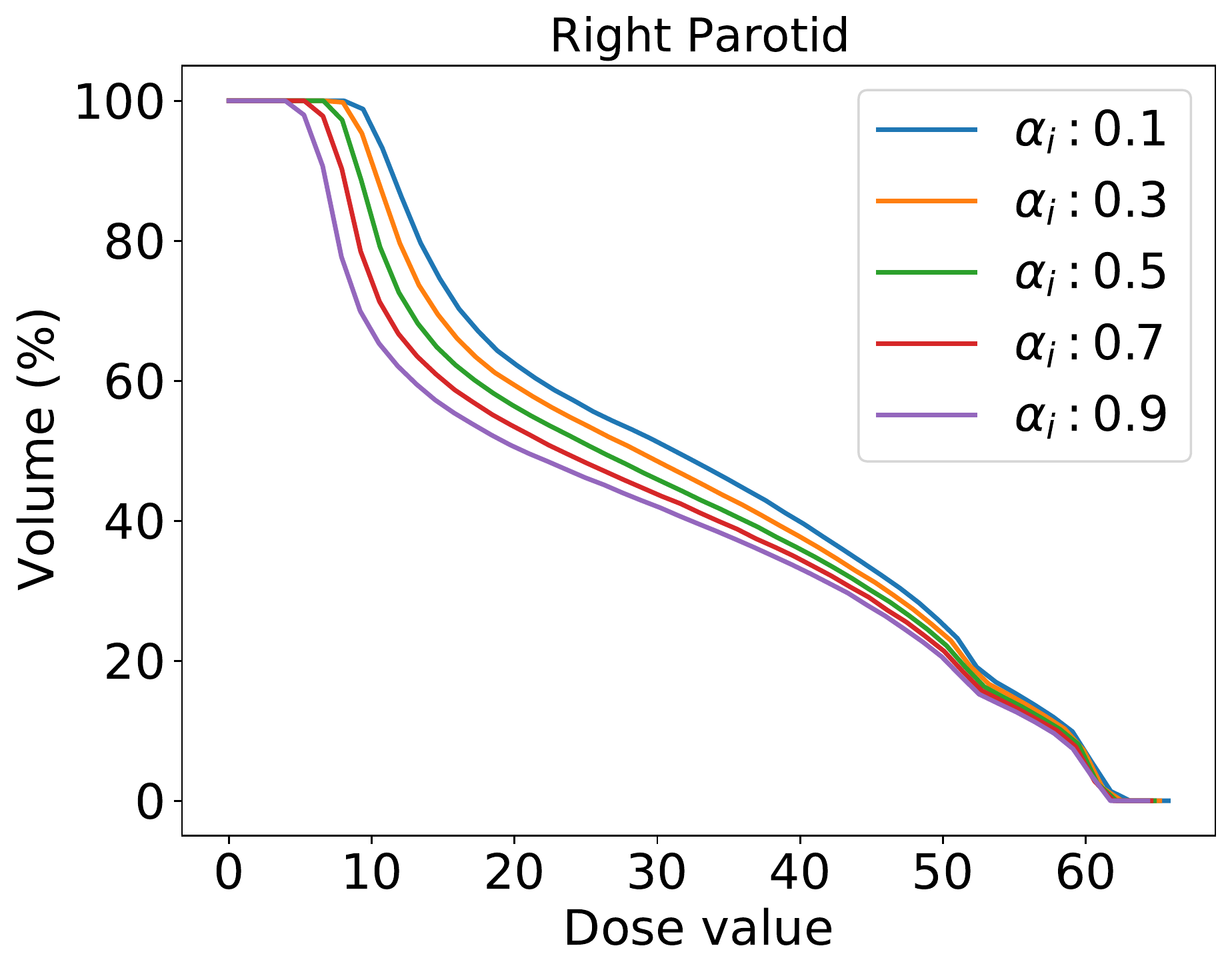}&
	\end{tabular}
	\caption{DVH of MMD with varying $\alpha_i$ values (Eq. \eqref{eq:mdd}) for {\bf (a)} brainstem;  {\bf (b)} spinal cord; {\bf (c)} larynx; {\bf (d)} left parotid; {\bf (e)} right parotid.}
	\label{fig:diff_alpha}
\end{figure}

Figure \ref{fig:dvh_3_models} superimposes the DVHs for the right parotid organ from the lower-extreme, upper-extreme and base models for a randomly chosen test patient.  This figure demonstrates the range obtainable is going between the lower- and upper-extreme models.  The model predicted range is a reflection of the range in the training set.  Further, we  observe that the DVH of the lower-extreme model is almost always below that of the base model, and also that the base model DVH is almost always below that of the upper-extreme model. In   figure \ref{fig:diff_alpha}, the DVH for the model dose distribution ($\text{MDD}^{{OAR}_{i}}$) of different OARs are plotted as a function of $\alpha_i$ (see Eq. \eqref{eq:mdd}). We observe that the higher the $\alpha_i$, the lower the proposed dose values, as expected.  Table \ref{tab:dose_range} shows the dose range between the lower-extreme and upper-extreme models for the test patients. The mean dose ranges for OARs show that a clinically reasonable range is achieved, reflective of the dose range in the training set patients.

\begin{table}[h]
	\small
	\centering
	\begin{adjustbox}{width=1\textwidth}
		\begin{tabular}{lcccccc}
			\hline
			& brainstem (max              &  \multicolumn{1}{c}{spinal cord (max} & \multicolumn{1}{c}{larynx (mean} & \multicolumn{1}{c}{left parotid (mean} & \multicolumn{1}{c}{right parotid (mean}\\
			    &  dose)&   dose)&   dose)&  dose)&  dose) \\ \hline
			25\% range (Gy)   &  14.15&   14.28&   5.11&  4.61&  5.23 \\
			75\% range (Gy)   &  19.93&   15.55&   6.79&  8.80&  8.24 \\
			mean range (Gy)   & 16.14 $\pm$ 1.09  & 14.75 $\pm$ 1.81 &  5.76 $\pm$ 1.05  &  6.08 $\pm$ 1.98 &  6.57 $\pm$ 1.45 \\ \hline
		\end{tabular}
	\end{adjustbox}
	\caption{\small OAR dose range (Gy) between lower-extreme and upper-extreme models in test patients.}
	\label{tab:dose_range}
\end{table}

\begin{figure}[hh!!]
	\centering
	\begin{tabular}{ll}
		\includegraphics[width=7cm]{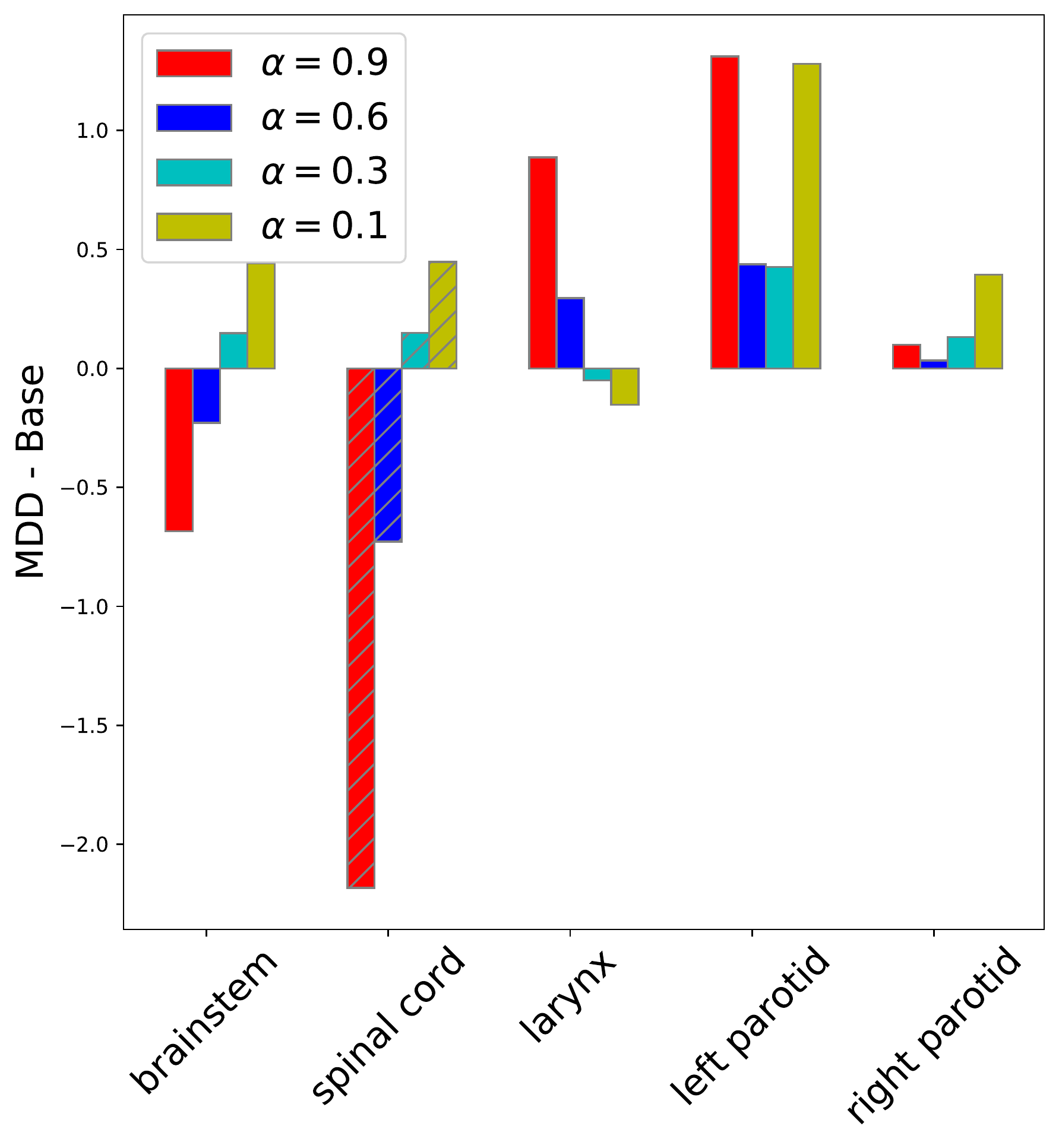}&	\includegraphics[width=7cm]{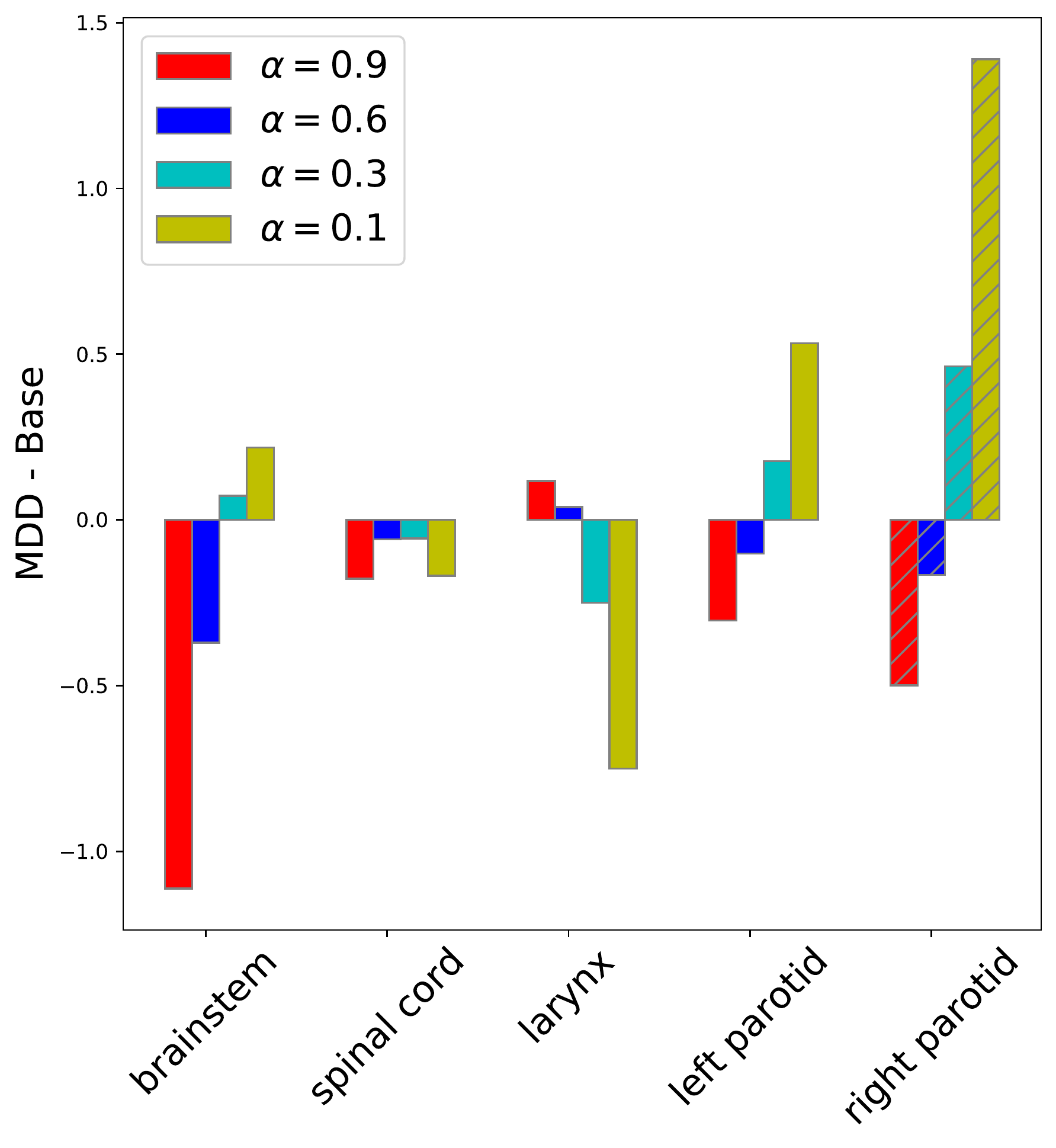}\\
	\end{tabular}
	\caption{Trade-off plot of MDD dose decrease to one OAR (from increasing risk parameter $\alpha_i$) leading to collateral dose increase to other OARs:  {\bf(a)} dose decrease to spinal cord below the base model dose is accompanied by dose increase to larynx above the base model dose; {\bf(b)} dose decrease to right parotid below the base model dose is accompanied by dose increase to larynx above the base model dose.}
	\label{fig:trade_offs}
\end{figure}

Next, we explore the trade-off effects of $\text{MDD}^{{OAR}_{i}}$ extreme models. We set different risk parameter values for $\alpha_i\in\{0.1,0.3,0.6,0.9\}$ in Eq. \eqref{eq:mdd} for a specific OAR and examine the effect on that specific OAR as well as the trade-off effect on other OARs. The bar plot in figure \ref{fig:trade_offs} shows the $\texttt{Deviation} = \texttt{MDD}^{{OAR}_{i}}-\texttt{Base model dose}$ for two scenarios: (a) spinal cord risk parameter values are varied; and (b) right parotid risk parameter values are varied.  In figure \ref{fig:trade_offs}(a), as expected, the spinal cord dose decreases below the base model dose with increasing $\alpha_i$ (bars filled with dashed lines), but leads to trade-off dose increase to the larynx above the Base model dose. In figure \ref{fig:trade_offs}(b), right parotid dose decreases below the base model dose with increasing $\alpha_i$ (bars filled with dashed lines), which is accompanied by a trade-off dose increase above the Base model dose to larynx, but also accompanied by synergistic dose decreases to brainstem and left parotid.

\subsection{Adjustable Plan}
\begin{figure}[hh!!]
	\centering
	\begin{tabular}{ll}
		\includegraphics[width=7.5cm]{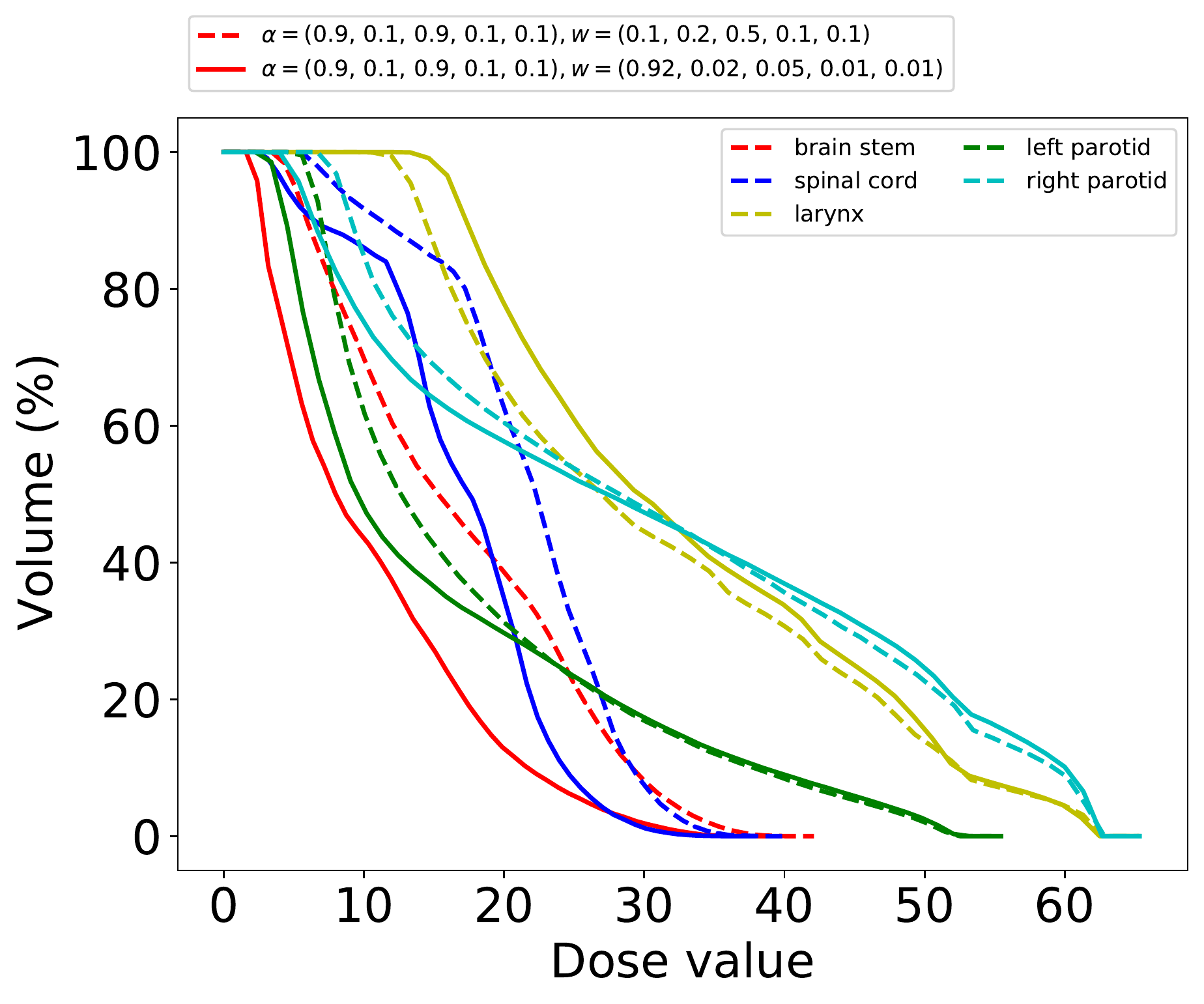}&	\includegraphics[width=7.5cm]{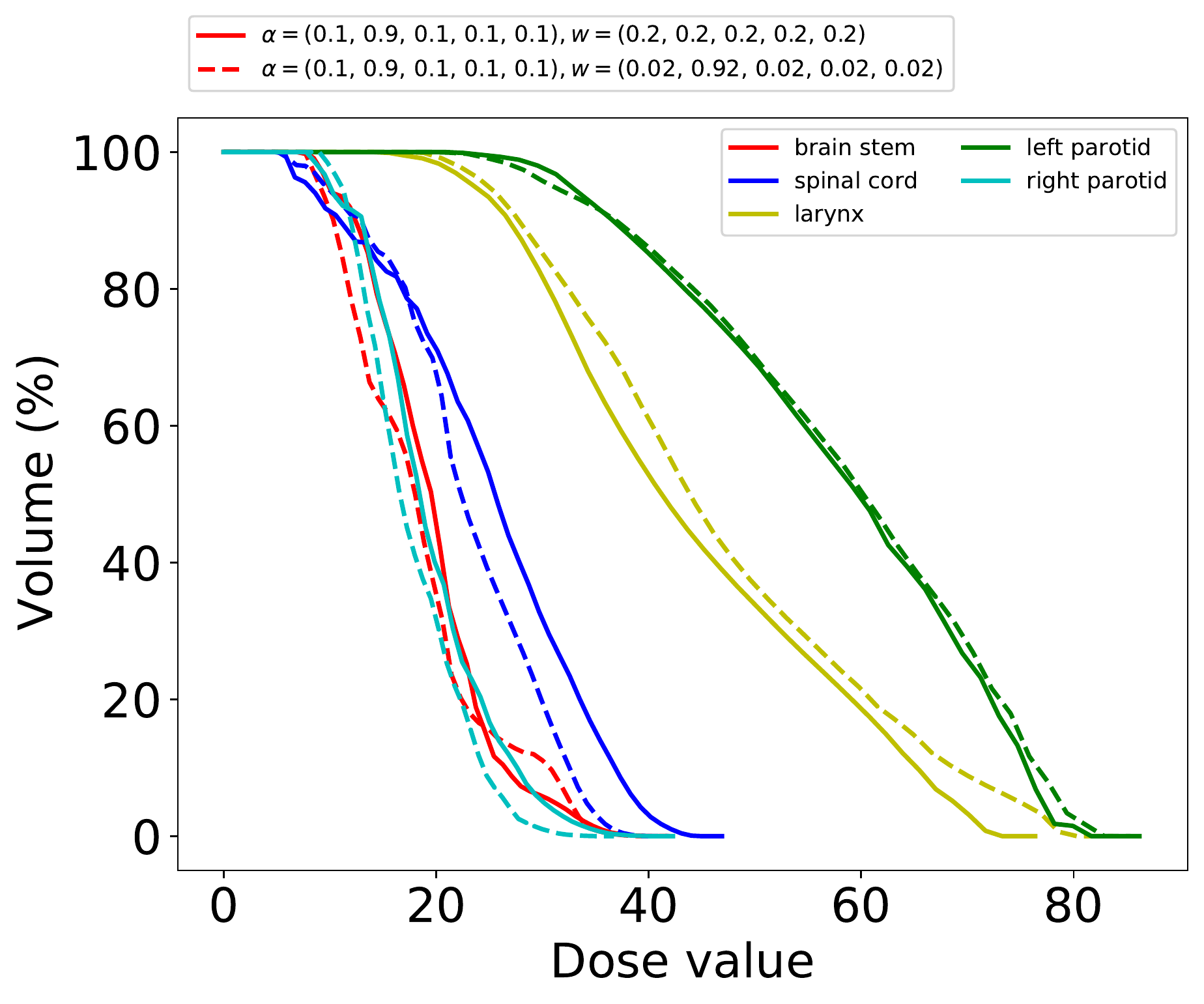}\\
	\end{tabular}
	\caption{{\bf(a)} Patient "A": DVHs for two different $\alpha_i$ and $w_i$ combinations to vary brainstem dose. Red lines indicate DVHs for brainstem for the two combinations.  {\bf(b)} Patient “B”: DVHs for two different $\alpha_i$ and $w_i$ combinations to vary spinal cord dose. Spinal cord DVHs for the two combinations are shown as blue lines.}
	\label{fig:single}
\end{figure}

The adjustable plan, defined in \eqref{eq:final_plan}, allows us to examine adjustable dose distributions (and DVHs) obtained by varying the risk parameters and organ weights in real-time. Figure \ref{fig:single} illustrates the DVH plot for two different test patients with different combinations of $\alpha_i$ and $w_i$ (the entries of vectors $\alpha_i$ and $w_i$ in the figure are in the order: brainstem, spinal cord, larynx, left parotid, right parotid). For patient “A”, in figure  \ref{fig:single} (a), it is assumed that low brainstem dose is to be prioritized, thus $\alpha_1$, which is associated with this OAR, is set to a high value. Further, we examined two different  organ weights $w_1$ for this organ (see dashed lines vs solid lines). We can see that as we increase $w_1$, the adjustable plan proposes lower dose to brainstem but, at the same time, larynx receives higher dose, indicating a trade-off. For patient “B”, in figure \ref{fig:single} (b), it is assumed that low spinal cord dose is to be prioritized. Hence, $\alpha_2$ is set to $0.9$. Similarly, different importance weight values $w_2$ are tested. This plot shows that the higher $w_2$ is set, the lower the dose to spinal cord. 

\section{Discussion}
This work addresses a drawback of the traditional Machine Learning prediction of a patient’s radiation therapy dose distribution, which is that a single dose distribution output in 3-dimensions may not be satisfactory.  The reason is that the output dose distribution is a Machine Learning “average” over many potential dose distributions in the training set.  The variation in dose distributions in the training set stems from both planner and clinician-driven trade-offs in doses between OARs, a consequence of the variability in the planning trajectory and the clinician’s desire to increase sparing to selective OARs based on patient-specific clinical parameters.  Thus, a single Machine Learning dose distribution output for a patient leaves the planner and clinician with little room to further personalize the dose distribution to that specific patient.

This drawback has been addressed in earlier work that used Pareto solutions \cite{nguyen2019generating, bohara2020using} in the Machine Learning training set to exploit a range of solutions.  However, doing so requires the generation of a large number of Pareto solutions, and, importantly, these Pareto solutions are not actual clinical plans generated by an expert human planner.  Here, instead, we recognize that the training set of clinical plans inherently includes dose trade-offs between OARs as a result of planner variability and patient-specific clinician preference. We exploit this variability to offer adjustable Machine Learning dose distributions that can be created in real-time to preferentially spare desired OARs (with concomitant dose increases to other OARs). An advantage of this approach is that the OAR dose range available in the Machine Learning solutions is a direct reflection of the dose range in the training set. Since the dose range in the training set is from clinically acceptable treatment plans, the Machine Learning solutions, in turn, are expected to stay within this clinically acceptable range.

There are several limitations to the implementation of this work.  The number of patients in the dataset is small, and all patients are from a single institution.  Moreover, the range of OAR doses available in the training set (and hence the range available in the Machine Learning model) is dependent, to some extent, on the idiosyncrasies of the planning procedure and clinician preference at the institution.  Our future goal is to check the generalizability of the Machine Learning adjustable dose distributions developed here to other institutions.

\section{Conclusions}
We present a Machine Learning framework for proposing adjustable head-and-neck cancer radiation therapy dose distributions. For a given patient, the adjustable solution presented here allows planners and clinicians the ability to prioritize and increase dose sparing to selected organs-at-risk in real-time. In contrast, traditional Machine Learning formulations only present a single dose distribution solution with no opportunity for the planner/clinician to make changes, potentially rendering the solution clinically unsatisfactory.  The adjustable solution is shown to offer clinically meaningful latitude in the ability to trade-off doses between organs-at-risk.

\newpage
{
	\bibliographystyle{plain}
	\bibliography{ref}
}

\end{document}